\documentclass[%
twocolumn,
showpacs,preprintnumbers,
 amsmath,amssymb,
 aps,
 prl,
floatfix,
]{revtex4-1}

\usepackage{soul,color}
\usepackage{natbib}

\usepackage[section]{placeins}
\usepackage{float}
\usepackage{graphicx}
\usepackage{stackengine}
\usepackage{xcolor}
\usepackage{dcolumn}
\usepackage{bm}
\DeclareMathOperator{\Tr}{Tr}
\usepackage{hyperref}

\begin{document}

\title{Fluorescence spectrum and thermalization in a driven coupled cavity array}

\author{Dainius Kilda}
\affiliation{SUPA, School of Physics and Astronomy, University of St Andrews, St Andrews, KY16 9SS, United Kingdom}
\author{Jonathan Keeling}%
\affiliation{SUPA, School of Physics and Astronomy, University of St Andrews, St Andrews, KY16 9SS, United Kingdom}

\date{\today}

\begin{abstract}
We calculate the fluorescence spectra of a driven lattice of coupled cavities.  To do this, we extend methods of evaluating two-time correlations in infinite lattices to open quantum systems;  this allows access to momentum resolved fluorescence spectrum. We illustrate this for a driven-dissipative  transverse field anisotropic XY model. By studying the fluctuation dissipation theorem, we find the emergence of a quasi-thermalized  steady state with a temperature dependent on system parameters; for blue detuned driving, we show this effective temperature is negative. In the low excitation density limit, we compare these numerical results to analytical spin-wave theory, providing an understanding of the form of the distribution function and the origin of quasi-thermalization.
\end{abstract}

\maketitle


By driving a system out of equilibrium, it is possible to stabilize states of matter that are either not known or are hard to achieve in thermal equilibrium.  Classically, driven systems have been extensively studied in the framework of pattern formation and dynamics~\cite{cross_hohenberg93}.  The study of quantum systems driven far from equilibrium is currently very active, in fields ranging from ultracold atoms~\cite{bloch2008many, ritsch2013cold, daley2014quantum, langen2015ultracold} to optically induced superconductivity~\cite{mitrano16}, and hybrid matter-light systems~\cite{carusotto2013quantum}. One such class of system is driven dissipative lattices~\cite{schmidt2013circuit, noh2016quantum, hartmann16}.
This is motivated by a variety of experimental platforms, including photonic crystal devices with quantum dots~\cite{majumdar2012cavity}, micropillar structures in semiconductor microcavities~\cite{sala2015vg}, trapped ions~\cite{toyoda2013experimental}, and microwave cavities and superconducting qubits~\cite{houck2012chip,fitzpatrick17}. 
Depending on the combination of couplings and driving used, many different models can be realized, and for many of these models, driving and dissipation allow one to induce a wider variety of 
collective states than occur in thermal equilibrium~\cite{carusotto2009fermionized, hartmann2010polariton, grujic2012non,nissen2012nonequilibrium,joshi2013quantum, jin2013photon, jin2014steady, biella2015photon,schiro2016exotic,lee2013unconventional}.

Most theoretical work on driven-dissipative lattices has focused on using order parameters or equal-time correlation functions to identify the phase diagram.  For coherent driving, such observables correspond to measuring the elastically scattered light. Less attention has been paid to the properties of the incoherent fluorescence from such lattices. From the quantum optics perspective, incoherent fluorescence of a coherently driven system  can reveal interactions and coherence times, as known for the Mollow triplet fluorescence~\cite{breuer2002theory}, which has been seen in candidate systems for coupled cavity arrays such as quantum dots~\cite{vamivakas2009spin} and superconducting qubits coupled to microwave cavities~\cite{lang11rf}. In extended systems, one can also access momentum-resolved spectra, e.g.  by measuring the interference of light emitted from different cavities. Moreover, second order correlations distinguish bunching or antibunching of photons --- as studied theoretically for a pair of coupled cavities~\cite{liew10single,bamba11single}. Applied to extended systems, such measurements can make contact with quantities typically seen with condensed matter probes such as angle resolved photon emission, spectroscopic scanning tunneling microscopy, or neutron scattering.  i.e. they measure the excitation and fluctuation spectrum of a correlated state, revealing the nature of correlated states.

There are other reasons to anticipate that calculations of two-point and two-time correlations can provide understanding beyond single-time observables.  Firstly, for any correct treatment of a finite size system, symmetry breaking should not occur.  This can also be true for certain numerical approaches in infinite systems: unless one uses the non-commuting limits of symmetry-breaking fields and system size,  one finds a steady state density matrix with equal mixtures of symmetry-broken states~\cite{rodriguez17hysteresis,fink2017signatures}.   Two-time correlations allow one to instead ask  how long symmetry breaking  persists in response to a probe --- i.e., long time correlations correspond to divergences of the zero frequency response of a system.   For driven systems, similar results may be extended to the treatment of limit cycles and `non-equilibrium time crystals'~\cite{chan2015limit,schiro2016exotic,iemini2017boundary}. The density matrix, as an ensemble averaged quantity,  involves averaging over the phase (or equivalently origin in time) of any limit cycle, washing out any time dependence in the density matrix.  In contrast two-time correlations  reveal such cycles as a diverging response at non-zero frequency.

Another motivation for studying two-time correlations is to investigate thermalization.  Thermalization in driven systems has been studied in a number of contexts, including the  `low energy effective temperature' in the Keldysh field theory of driven atom-photon systems~\cite{diehl2008quantum, diehl2010dynamical,  oztop2012excitations, dalla2013keldysh, buchhold2013dicke, sieberer2013dynamical, sieberer2016keldysh} and the mode populations in photon~\cite{klaers2010bose,kirton2015thermalization} and polariton condensates~\cite{doan05,kasprzak08, sun17thermalisation}.  This steady state behavior in a continuously driven system can also be connected to the emergence of a prethermalized state following a sudden quench in an isolated system~\cite{polkovnikov2011colloquium,eisert2015quantum,langen16pretherm} --- in such a prethermalized state, there is a flow of energy between degrees of freedom at different  scales.  For a thermalized state, we expect the density matrix takes the Gibbs form
\begin{math} 
\rho = \exp\left(-H_{\text{eff}}/T_{\text{eff}}\right)
\end{math}
with  some effective Hamiltonian $H_{\text{eff}}$.   One may note however that \emph{any} density matrix can be written in the Gibbs form;   to make the criterion meaningful one thus needs a method to independently determine $H_{\text{eff}}$.  This means simultaneously measuring the occupations and densities of available states --- this is the essence of the fluctuation dissipation theorem, which we discuss below.

In this letter, we find the two-time correlations of a driven dissipative  lattice, and see the emergence of a quasi-thermalized state.  We calculate both on-site and inter-site correlations, giving access to the momentum resolved fluorescence spectrum of a driven coupled cavity array. In order to eliminate boundary and finite-size effects, we work always with the translationally invariant infinite lattice. On-site calculations in a finite-size lattice have also been recently studied for the XXZ model~\cite{wolff2018evolution}. While the methods we present are general, our work will focus on the transverse field anisotropic XY model (which has both the Ising and XY models as special cases), a driven dissipative realization of which was proposed by~\citet{bardyn2012majorana}, and the steady state properties  studied~\cite{joshi2013quantum,mascarenhas2015matrix} using matrix product state approaches. As shown in~\cite{bardyn2012majorana} and reviewed in the supplementary material~\cite{supplementary}, this model can be realized by an array of coupled cavities in the photon blockade regime, with a two-photon pump that creates pairs of photons in adjacent sites (see Fig.~\ref{CCA-Cartoon}).

\begin{figure}[hptb]
\includegraphics[width=0.60\linewidth]{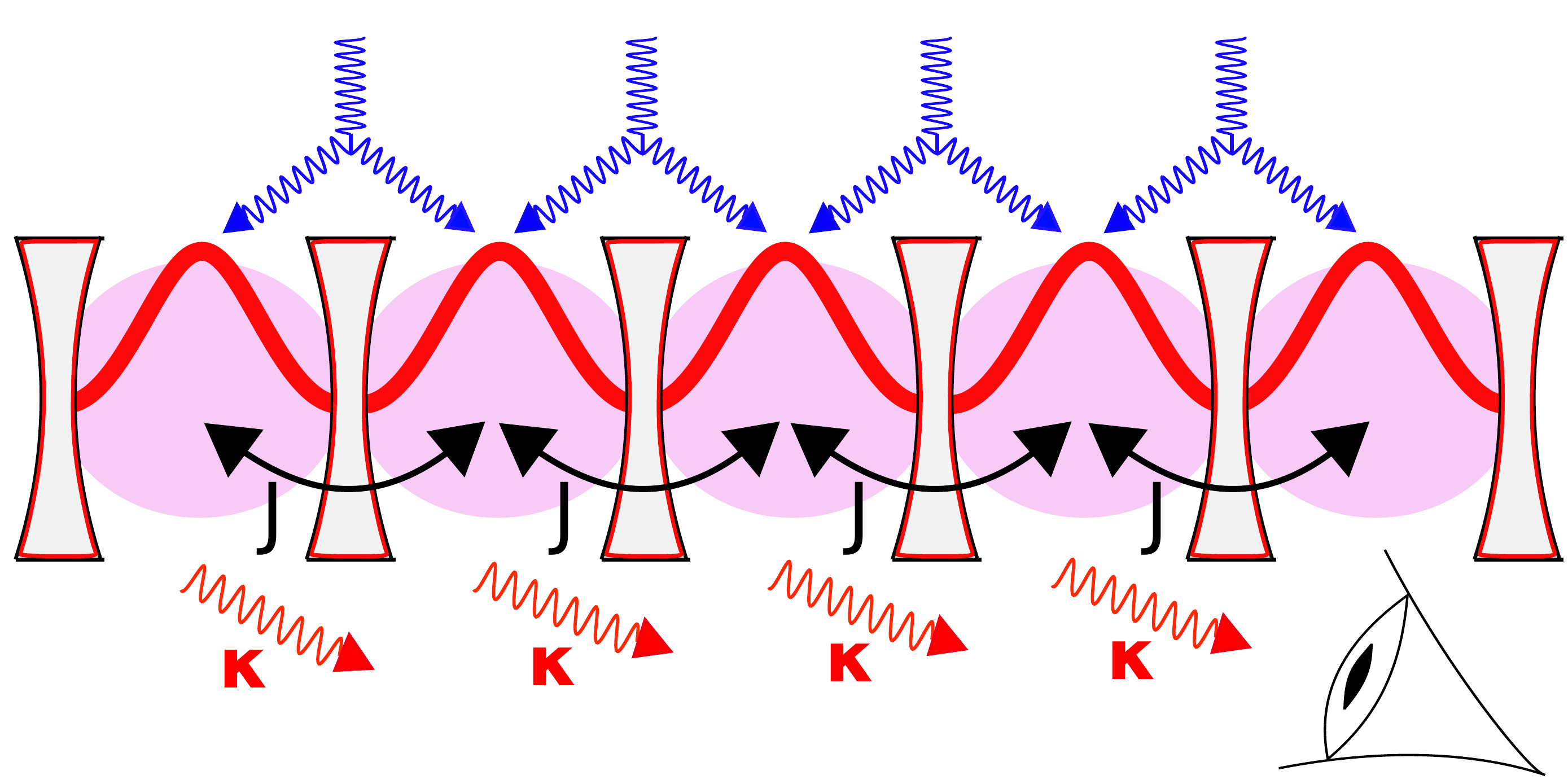}
\caption{Coupled cavity array with hopping $J$, photon loss $\kappa$ and two-photon pumping (blue line). When strong nonlinearity (purple shading) in each cavity leads to photon blockade, this yields the transverse field anisotropic XY model~\cite{bardyn2012majorana}.}
\label{CCA-Cartoon}
\end{figure}

Following~\cite{joshi2013quantum,supplementary}, working
in the rotating frame of the pump, the effective Hamiltonian has the form:
\begin{math} 
  H = - J \sum_j \left[ g \sigma_j^z + \frac{1+\Delta}{2} \sigma_j^x \sigma_{j+1}^x + \frac{1-\Delta}{2} \sigma_j^y \sigma_{j+1}^y \right].
\end{math}
The dimensionless transverse field $g$ depends on the pump-cavity detuning, and the anistropy parameter $\Delta$, given by the ratio of pump strength and photon hopping $J$.  For $\Delta = 1$ we recover the Ising model and for $\Delta = 0$  the isotropic XY model. In the following we will work in units of J. For the driven system, the Hamiltonian is accompanied by photon loss at rate $\kappa$ into  empty radiation modes.  We thus have the master equation:
\begin{multline} \label{ME}
\; \; \; \; \; \; \partial_t \rho = \mathcal{L}\lbrace \rho \rbrace = -i\left[H,\rho \right] \\+ \frac{\kappa}{2} \sum_j \left(2 \sigma_j^{-} \, \rho \sigma_j^{+} - \sigma_j^{+} \sigma_j^{-} \, \rho - \rho \, \sigma_j^{+} \sigma_j^{-} \right).
\end{multline}
While a non-driven system would equilibrate with the bath, the  time-dependent driving breaks detailed balance and leads instead to a  nonequilibrium steady state (NESS).

The fluctuation and response spectra discussed above require evaluating
two-time correlation functions which, for a Markovian system, can
be found using the quantum regression theorem~\cite{breuer2002theory}:
\begin{equation} \label{Q-reg}
\left< O_2^{(j)}(t) O_1^{(i)}(0) \right> = \Tr\left[O_2^{(j)} \, e^{t \mathcal{L}} \, O_1^{(i)}\rho_{\text{ss}}\right],
\end{equation}
where $i,j$ label two lattice sites and $1,2$ two local operators.
In order to compute this for an infinite lattice, we employ matrix product state (MPS) methods. We first find the steady state  $\rho_{\text{ss}}$ of the master equation \eqref{ME}.  We do this by using the infinite Time Evolving Block Decimation (iTEBD) algorithm~\cite{schollwock2011density,orus2014practical} 
to find the translationally invariant infinite MPS such that $\mathcal{L}\left\{ \rho_{\text{ss}} \right\}=0$.  Starting from the NESS, we then calculate two-time correlations using Eq.~\eqref{Q-reg}.
Because applying local operators $\hat{O}_{1}$  to $\rho_{\text{ss}}$ breaks  translational invariance, we can no longer propagate using iTEBD.  
For a finite size lattice, TEBD could be used, but this restricts the extent of correlations in both space and time, as excitations are reflected from the boundaries~\cite{phien2012infinite}.
Fortunately, a method to find such correlations in an infinite lattice has been developed by~\citet{banuls2009matrix} for unitary evolution. This approach~\cite{banuls2009matrix}, which we extend to open systems, writes the time evolution between applying $\hat{O}_1$ and $\hat{O}_2$ as a tensor network, and contracting this network gives the desired correlator (see~\cite{supplementary} for details).  
 
Using this approach, we calculate the  fluctuation spectrum $S_{O, O^{\dag}}\left( \omega \right)$ and the response function of the system $\chi^{\prime\prime}_{O, O^{\dag}}\left( \omega \right)$
which are at the heart of the fluctuation-dissipation theorem~\cite{breuer2002theory, sieberer2016keldysh},
\begin{math} 
S_{O, O^{\dag}}\left( \omega \right) = F\left( \omega \right) \chi^{\prime\prime}_{O, O^{\dag}}\left( \omega \right),
\end{math} 
with the distribution function $F(\omega)$ discussed below.
Both $S_{O, O^{\dag}} \left( \omega \right)$ and $\chi^{\prime\prime}_{O, O^{\dag}} \left( \omega \right)$ are the Fourier transforms of two-time correlations 
\begin{align} \label{Corrs-def}
\tilde{S}_{O, O^{\dag}}\left( t \right) &= \frac{1}{2} \left< \lbrace \hat{O}\left( t \right) , \hat{O}^{\dag} \left( 0 \right) \rbrace  \right>,
\\
\tilde{\chi}_{O, O^{\dag}}\left( t \right) &= i \theta \left( t \right) \left< [\hat{O}\left( t \right) , \hat{O}^{\dag} \left( 0 \right)] \right>,
\end{align}
which we may evaluate using Eq.~(\ref{Q-reg}).

\fboxrule=0pt\relax
\fboxsep=0pt\relax
\begin{figure} 
  \includegraphics[trim={1.9cm 18.8cm 11.0cm 1.8cm},clip,width=1.0\linewidth]{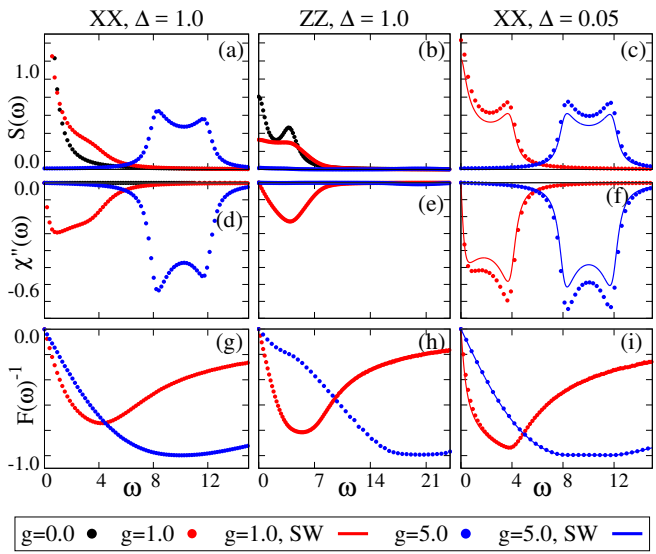}
  \caption{Spectrum of fluctuations $S(\omega)$, imaginary part of response
    function $\chi^{\prime\prime}(\omega)$, and inverse distribution
    function $F(\omega)^{-1}$.  Left two columns: Ising limit $\Delta=1$,
    Right column shows $\Delta=0.05$ where spin-wave theory (solid lines)
    matches well. Energies given in units of $J$. Other parameters used: $\kappa=0.5$.}
\label{1D-spectra}
\end{figure}

Figure~\ref{1D-spectra} shows the on-site ($i=j$) fluctuation and response functions in frequency domain for $\hat{O}_1=\hat{O}_2=\hat{O} \in \left\{ \sigma^x, \sigma^z \right\}$ and a range of values of transverse field $g$. We show both the Ising limit, ($\Delta=1$, left two columns)
as well as at small $\Delta$ (right column), where analytic results can be found using spin-wave theory as discussed further below.
The panels (a--c) show $S\left(\omega\right)$ which measures the occupations while, panels (d--f) show response function $\chi^{\prime\prime}\left( \omega \right)$, which measures the density of states (DoS).  We note that while at $g=0,1$ we see $S\left(\omega\right)$ for $\sigma^x$ is peaked at $\omega=0$, its value always remains finite as there is no phase transition in this open one-dimensional system~\cite{lee2013unconventional,joshi2013quantum}.  As we will discuss later, the form of the density of states seen here can be understood from the momentum resolved correlation functions.

The bottom row of Fig.~\ref{1D-spectra} shows the inverse distribution  functions $F(\omega)^{-1}=\chi^{\prime\prime}_{O, O^{\dag}}\left( \omega \right)/S_{O, O^{\dag}}\left( \omega \right)$ for $\hat{O}=\sigma^x, \sigma^z$
respectively. In an equilibrium system, the distribution function $F\left(\omega \right)$
 depends only on whether $\hat{O}$ obeys Fermionic or
Bosonic (anti-)commutation relations; for Bosons it is:
\begin{math} 
F\left(\omega\right) \equiv 2 n_B\left(\omega\right) + 1 = \coth((\omega-\mu)/2T).
\end{math}
In a driven dissipative system, $F\left(\omega\right)$ may take a more general form. However as identified in other contexts~\cite{diehl2010dynamical, diehl2008quantum, oztop2012excitations, sieberer2016keldysh, carusotto2013quantum, dalla2013keldysh, buchhold2013dicke, sieberer2013dynamical}, quasi-thermalisation of low energy modes often occurs, leading to the identification of a low energy effective temperature
$F(\omega) \sim {2T_{\text{eff}}}/{\omega}$. Note that since all calculations are performed in the rotating frame, all frequencies are measured relative to the pump frequency --- i.e. the pump frequency acts as an effective chemical potential $\mu$ that sets the frequency at which $F(\omega)$ diverges.

As seen in Fig.~\ref{1D-spectra}(g,h), $F(\omega)^{-1}$ is linear $\omega \to 0$ indicating the emergence of a low energy effective temperature in this model.
Because the power spectrum of physical operators is positive, there is a minimum possible fluctuation contribution for a given dissipation, meaning $|F(\omega)|^{-1} \leq 1$.  At high frequencies the distribution function of a fully thermalised system asymptotically approaches this value.
In our non-equilibrium system we see that in some cases the inverse distribution $|F(\omega)|^{-1}$ approaches $1$ over a range of frequencies, however in all cases it falls falls below one at higher frequencies, indicating higher fluctuations than for a thermal state. The results shown give some indication that, at least for Fig.~\ref{1D-spectra}(g), the $F(\omega)$ approaches a thermal form more closely at larger $g$.

\begin{figure} 
  \includegraphics[trim={2.2cm 23.38cm 11.0cm 1.8cm},clip,width=1.0\linewidth]{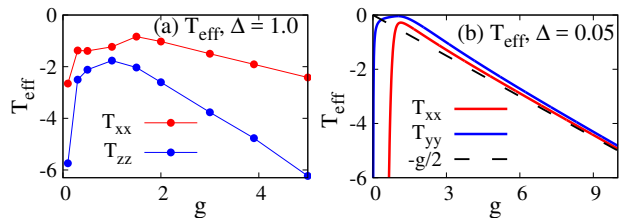}
  \caption{The effective temperature $T_{\text{eff}}$ against transverse field $g$. We find $T_{\text{eff}}$ by fitting $F(\omega) \simeq A \coth(b \omega)$ for low frequencies ($\omega \leq 1.0$), and plotting $T_{\text{eff}} \equiv A/2b$.    (a) MPS results for $\sigma^{x,z}$ fluctuations and the transverse-field Ising limit ($\Delta=1.0$); (b) Spin wave results for $\sigma^{x,y}$ fluctuations at $\Delta = 0.05$. Energies given in units of $J$. Other parameters used: $\kappa=0.5$.}
\label{SW-data}
\end{figure}

The right column of Fig.~\ref{1D-spectra} compares the MPS results (points) to analytic spin-wave theory~\cite{mattis2006theory,supplementary}, which is valid if the density of excitations is small.  We see a good agreement between spin-wave theory and MPS numerics at $\Delta=0.05$ for $\sigma^x$ correlations (we do not show the $\sigma^z$ spectra for this $\Delta$, as these vanish in the linearised spin-wave theory). Remarkably, the agreement for $F(\omega)$ is better than for $S\left(\omega\right), \chi^{\prime\prime}\left( \omega \right)$ individually.
It is notable that despite being a linear (i.e. non-interacting) theory, the spin-wave result reproduces both the low energy effective temperature and the emergent plateau $F(\omega) \simeq 1$ at intermediate frequencies.  The distribution function of spin-wave theory can be understood as a weighted average of $k$-dependent function $F(\omega,k) = (2 T_{\text{eff},k} +\lambda_k \omega^2)/\omega$, with weighting by the $k$-dependent density of states~\cite{supplementary}. This form (which follows directly from the structure of the relevant linearised theory) leads directly to the existence of a low energy effective temperature.  The plateau at $F(\omega) \simeq 1$, seen only at larger $g$, results from the local spectra averaging over many momentum states~\cite{supplementary}, however the form $F(\omega,k)$ inevitably leads to $F(\omega) \propto \omega$ at high frequencies, corresponding to the breakdown of the plateau.



As well as the deviation from the thermal $F(\omega)$, a second distinction from an equilibrated system is that both the distribution and the low-energy effective temperature extracted differ depending on the system operator considered. Figure~\ref{SW-data}(a) shows how $T_{\text{eff}}$ of $\sigma^x$ and $\sigma^z$ correlators vary with transverse field $g$.
Fig.~\ref{SW-data}(b) shows similar results for the spin-wave theory at small $\Delta$ for $\sigma^x$ and $\sigma^y$ correlators (as noted above,
$\sigma^z$ correlators vanish in a linearised theory). We observe that for $\Delta \to 0$, $g \to \infty$ the $\sigma^{x,y}$ excitations thermalize to the same  effective temperature, $T_{\text{eff}} \approx -g/2$.  This can be understood as $T_{\text{eff},k}$ becomes $k$ independent in this limit,
see~\cite{supplementary}. 
 

We only show results for $g>0$ in Fig.~\ref{1D-spectra}, since there exists a simple duality allowing us to relate the form of $S\left(\omega\right)$, $\chi^{\prime\prime}\left(\omega\right)$, $F\left(\omega\right)^{-1}$ for values $g$ and $-g$. This duality, discussed in~\cite{joshi2013quantum} arises because a combination of $g \mapsto -g$ and a $\pi$ rotation of the spin on every second site leads to
$H \mapsto -H$.  (A more general discussion of such dualities can be found
in~\cite{li2017:inversion}.) This duality means that on changing the sign of $g$, the state of the
system should correspond to reversing the sign of all energies.  We may then note that fluctuation and dissipation spectra show different parity; $\chi^{\prime\prime}\left(- \omega \right) = -\chi^{\prime\prime}\left( \omega \right) $, while $ S\left( -\omega \right) = S(\omega)$, and so $F(-\omega)=-F(\omega)$. As such, the energy sign reversal under $g \mapsto -g$  yields a sign change of the distribution function and effective temperature.  We find $g < 0$ gives positive temperatures, and $g > 0$ negative temperatures. This is consistent with the spatial ordering seen~\cite{joshi2013quantum}: for $T_{\text{eff}} < 0$ there is a high energy anti-ferromagnetic state.
A more intuitive understanding of this comes from
the fact that $g$ is proportional to the pump-cavity detuning, so that $g<0$ corresponds to a red-detuned pump and consequent cooling, while $g>0$ corresponds to blue detuning. Blue-detuned pumping is typically associated
to heating; here it does lead to energy accumulation, but this induces a negative temperature state, rather than high positive temperatures.
At $g=0$, the susceptibility $\chi^{\prime\prime}(\omega)$ vanishes, so
$F(\omega)^{-1}=0$ and the effective temperature diverges. 

\begin{figure} 
  \includegraphics[width=1.0\linewidth]{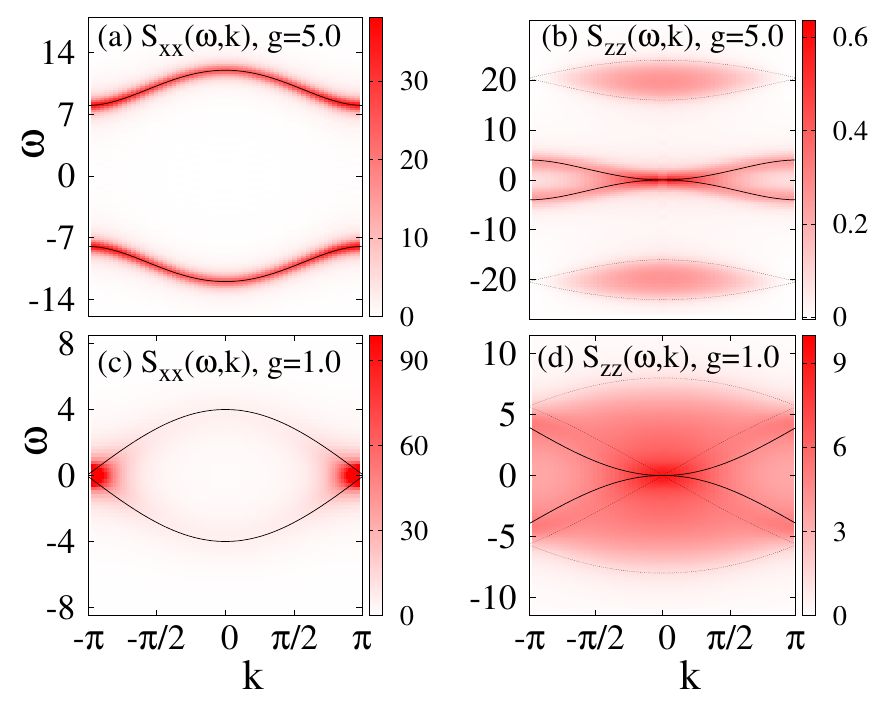}
\caption{$S(\omega, k)$ momentum-resolved fluctuation spectrum for excitations of: (a) $\sigma^x$ at $g=5.0$; (b) $\sigma^z$ at g=5.0; (c) $\sigma^x$ at $g=1.0$; (d) $\sigma^z$ at $g=1.0$. Energies given in units of $J$. Other parameters used: $\kappa=0.5$.}
\label{2D-spectra}
\end{figure}
So far we have evaluated correlation functions at equal positions; this corresponds to recording all light from one cavity, which implicitly integrates over momentum.  More information on the structure of the correlations is available if we consider the momentum-resolved spectrum.  This requires evaluating correlations at non-equal sites $i,j$, and performing a double Fourier transform with respect to separation in time $t$ and space $|i-j|$.  The resulting fluctuation spectra $S(\omega,k)$ are displayed in Fig.~\ref{2D-spectra} (The response function $\chi^{\prime\prime}(\omega,k)$ shows similar features as $S(\omega,k)$).  We show the case for $\hat{O}=\sigma^x, \sigma^z$, and two values of $g$ (we consider only $g>0$, since the duality discussed above allows one to understand the effects of a sign change of $g$).  All the features visible in these spectra can be described straightforwardly using excitation spectra derived from the Jordan-Wigner solution of $H_{\text{TFI}}$ (see e.g.~\cite{sachdev2007quantum} for details).

At large positive $g$, the NESS is known~\cite{joshi2013quantum} to be a
maximum energy state with spins pointing in the $-\hat{z}$ direction, opposing the magnetic field.  The spectrum of the $\sigma^x$ operator corresponds to single spin flips, so follows the single particle dispersion $\omega(k) = \epsilon(k) \equiv 2J\sqrt{1+g^2+2g\cos(k)}$ (where we consider \emph{de}-excitations of the \emph{maximum} energy state).  This expression is shown by the black line in Fig.~\ref{2D-spectra}(a).  In contrast, the $\sigma^z$ operator corresponds to two-particle excitations, which come in two varieties. The first one is a two-particle continuum with $\omega=\epsilon(k_1) + \epsilon(k_2), k_1 + k_2=k$.   The envelope of these states is given by $\epsilon_{\text{min}}(k) < \omega(k) < \epsilon_{\text{max}}(k)$ with $\epsilon_{\text{max}/\text{min}}(k) = 4J\sqrt{1 + g^2 \pm 2g\cos(k/2)}$, shown by the dotted black lines in Fig.~\ref{2D-spectra}(b).  The other kind of excitations involves  scattering existing particles from mode $q$ to $q+k$, i.e. $\omega(k) = \Delta\epsilon(q,k) \equiv \epsilon(q+k)-\epsilon(q)$. The dominant contribution comes from $q=0$, since this corresponds to the maximum energy mode, which is maximally occupied for a negative temperature state.
The black solid line shows $\Delta\epsilon(0,k)$ which indeed matches the dominant feature observed. Given these momentum resolved results, the momentum integrated spectral functions in Fig.~\ref{1D-spectra}(a,c) can be easily understood, with peaks arising from van Hove singularities at the band edges.

Near $g=1.0$ the NESS  instead shows antiferromagnetic correlations.   The spectra here retain key features but are distorted.  In the $\sigma^x$ spectrum, Fig.~\ref{2D-spectra}(c) and Fig.~\ref{1D-spectra}(a,c), the peaks at $k=\pm\pi$ become  dominant. In the $\sigma^z$ spectrum, Fig.~\ref{2D-spectra}(d) and Fig.~\ref{1D-spectra}(b,d), the  scattering band and  two-particle continuum overlap. The black lines show the same expressions as discussed above.  A ground state phase transition occurs for $|g|<1$, hence the gap closing at $g=1.0$. In contrast, the NESS at $g=1.0$ already enters an antiferromagnetic state. As such, it is unsurprising these dispersions (which use normal state Jordan Wigner forms) do not match the spectrum as well as they did at $g=1.0$. As one continues to decrease $g \to 0$ the spectrum becomes further dominated by the modes near $\omega=0$, as seen in Fig.~\ref{1D-spectra}(a--d).

In conclusion, we have calculated the two-time correlations of a driven-dissipative coupled-cavity array, providing the fluorescence and absorption spectra. Due to the duality between red and blue detuned scenarios, we find that a blue pump-cavity detuning produces a quasi-thermalized state with a negative temperature. We have also shown how the structure of $F(\omega)$ and emergent thermalization can be understood using a spin-wave theory, and how momentum resolved fluorescence reveals the nature of quasiparticle excitations in the quasi-thermal state. The system we have studied here is in the photon blockade regime, with at most one excitation per site.  This restricts us to study ``first-order'' correlation functions.  When generalizing to problems with a larger on-site Hilbert space, second-order photon counting correlations may also be of interest, in revealing the coherence and statistics of any ordered state.  Our results illustrate how calculating such correlations of the fluorescence can provide new insights into the state
of many-body driven dissipative systems.

\begin{acknowledgments}
  DK acknowledges support from the EPSRC CM-CDT (EP/L015110/1).
  JK\ acknowledges support from EPSRC program “TOPNES” (EP/I031014/1).
  We are grateful to M. Hartmann and S. H. Simon for helpful discussions,
  and to M. Hartmann for helpful comments on the manuscript.
\end{acknowledgments}

\bibliography{TwoTimeCorrelationsCavityArray}

\begin{thebibliography}{64}%
\makeatletter
\providecommand \@ifxundefined [1]{%
 \@ifx{#1\undefined}
}%
\providecommand \@ifnum [1]{%
 \ifnum #1\expandafter \@firstoftwo
 \else \expandafter \@secondoftwo
 \fi
}%
\providecommand \@ifx [1]{%
 \ifx #1\expandafter \@firstoftwo
 \else \expandafter \@secondoftwo
 \fi
}%
\providecommand \natexlab [1]{#1}%
\providecommand \enquote  [1]{``#1''}%
\providecommand \bibnamefont  [1]{#1}%
\providecommand \bibfnamefont [1]{#1}%
\providecommand \citenamefont [1]{#1}%
\providecommand \href@noop [0]{\@secondoftwo}%
\providecommand \href [0]{\begingroup \@sanitize@url \@href}%
\providecommand \@href[1]{\@@startlink{#1}\@@href}%
\providecommand \@@href[1]{\endgroup#1\@@endlink}%
\providecommand \@sanitize@url [0]{\catcode `\\12\catcode `\$12\catcode
  `\&12\catcode `\#12\catcode `\^12\catcode `\_12\catcode `\%12\relax}%
\providecommand \@@startlink[1]{}%
\providecommand \@@endlink[0]{}%
\providecommand \url  [0]{\begingroup\@sanitize@url \@url }%
\providecommand \@url [1]{\endgroup\@href {#1}{\urlprefix }}%
\providecommand \urlprefix  [0]{URL }%
\providecommand \Eprint [0]{\href }%
\providecommand \doibase [0]{http://dx.doi.org/}%
\providecommand \selectlanguage [0]{\@gobble}%
\providecommand \bibinfo  [0]{\@secondoftwo}%
\providecommand \bibfield  [0]{\@secondoftwo}%
\providecommand \translation [1]{[#1]}%
\providecommand \BibitemOpen [0]{}%
\providecommand \bibitemStop [0]{}%
\providecommand \bibitemNoStop [0]{.\EOS\space}%
\providecommand \EOS [0]{\spacefactor3000\relax}%
\providecommand \BibitemShut  [1]{\csname bibitem#1\endcsname}%
\let\auto@bib@innerbib\@empty
\bibitem [{\citenamefont {Cross}\ and\ \citenamefont
  {Hohenberg}(1993)}]{cross_hohenberg93}%
  \BibitemOpen
  \bibfield  {author} {\bibinfo {author} {\bibfnamefont {M.~C.}\ \bibnamefont
  {Cross}}\ and\ \bibinfo {author} {\bibfnamefont {P.~C.}\ \bibnamefont
  {Hohenberg}},\ }\href {\doibase 10.1103/RevModPhys.65.851} {\bibfield
  {journal} {\bibinfo  {journal} {Rev. Mod. Phys.}\ }\textbf {\bibinfo {volume}
  {65}},\ \bibinfo {pages} {851} (\bibinfo {year} {1993})}\BibitemShut
  {NoStop}%
\bibitem [{\citenamefont {Bloch}\ \emph {et~al.}(2008)\citenamefont {Bloch},
  \citenamefont {Dalibard},\ and\ \citenamefont {Zwerger}}]{bloch2008many}%
  \BibitemOpen
  \bibfield  {author} {\bibinfo {author} {\bibfnamefont {I.}~\bibnamefont
  {Bloch}}, \bibinfo {author} {\bibfnamefont {J.}~\bibnamefont {Dalibard}}, \
  and\ \bibinfo {author} {\bibfnamefont {W.}~\bibnamefont {Zwerger}},\ }\href
  {\doibase 10.1103/RevModPhys.80.885} {\bibfield  {journal} {\bibinfo
  {journal} {Rev. Mod. Phys.}\ }\textbf {\bibinfo {volume} {80}},\ \bibinfo
  {pages} {885} (\bibinfo {year} {2008})}\BibitemShut {NoStop}%
\bibitem [{\citenamefont {Ritsch}\ \emph {et~al.}(2013)\citenamefont {Ritsch},
  \citenamefont {Domokos}, \citenamefont {Brennecke},\ and\ \citenamefont
  {Esslinger}}]{ritsch2013cold}%
  \BibitemOpen
  \bibfield  {author} {\bibinfo {author} {\bibfnamefont {H.}~\bibnamefont
  {Ritsch}}, \bibinfo {author} {\bibfnamefont {P.}~\bibnamefont {Domokos}},
  \bibinfo {author} {\bibfnamefont {F.}~\bibnamefont {Brennecke}}, \ and\
  \bibinfo {author} {\bibfnamefont {T.}~\bibnamefont {Esslinger}},\ }\href
  {\doibase 10.1103/RevModPhys.85.553} {\bibfield  {journal} {\bibinfo
  {journal} {Rev. Mod. Phys.}\ }\textbf {\bibinfo {volume} {85}},\ \bibinfo
  {pages} {553} (\bibinfo {year} {2013})}\BibitemShut {NoStop}%
\bibitem [{\citenamefont {Daley}(2014)}]{daley2014quantum}%
  \BibitemOpen
  \bibfield  {author} {\bibinfo {author} {\bibfnamefont {A.~J.}\ \bibnamefont
  {Daley}},\ }\href {\doibase 10.1080/00018732.2014.933502} {\bibfield
  {journal} {\bibinfo  {journal} {Adv. Phys.}\ }\textbf {\bibinfo {volume}
  {63}},\ \bibinfo {pages} {77} (\bibinfo {year} {2014})}\BibitemShut {NoStop}%
\bibitem [{\citenamefont {Langen}\ \emph {et~al.}(2015)\citenamefont {Langen},
  \citenamefont {Geiger},\ and\ \citenamefont
  {Schmiedmayer}}]{langen2015ultracold}%
  \BibitemOpen
  \bibfield  {author} {\bibinfo {author} {\bibfnamefont {T.}~\bibnamefont
  {Langen}}, \bibinfo {author} {\bibfnamefont {R.}~\bibnamefont {Geiger}}, \
  and\ \bibinfo {author} {\bibfnamefont {J.}~\bibnamefont {Schmiedmayer}},\
  }\href {\doibase 10.1146/annurev-conmatphys-031214-014548} {\bibfield
  {journal} {\bibinfo  {journal} {Annu. Rev. Condens. Matter Phys.}\ }\textbf
  {\bibinfo {volume} {6}},\ \bibinfo {pages} {201} (\bibinfo {year}
  {2015})}\BibitemShut {NoStop}%
\bibitem [{\citenamefont {Mitrano}\ \emph {et~al.}(2016)\citenamefont
  {Mitrano}, \citenamefont {Cantaluppi}, \citenamefont {Nicoletti},
  \citenamefont {Kaiser}, \citenamefont {Perucchi}, \citenamefont {Lupi},
  \citenamefont {Pietro}, \citenamefont {Pontiroli}, \citenamefont {Riccò},
  \citenamefont {Clark}, \citenamefont {Jaksch},\ and\ \citenamefont
  {Cavalleri}}]{mitrano16}%
  \BibitemOpen
  \bibfield  {author} {\bibinfo {author} {\bibfnamefont {M.}~\bibnamefont
  {Mitrano}}, \bibinfo {author} {\bibfnamefont {A.}~\bibnamefont {Cantaluppi}},
  \bibinfo {author} {\bibfnamefont {D.}~\bibnamefont {Nicoletti}}, \bibinfo
  {author} {\bibfnamefont {S.}~\bibnamefont {Kaiser}}, \bibinfo {author}
  {\bibfnamefont {A.}~\bibnamefont {Perucchi}}, \bibinfo {author}
  {\bibfnamefont {S.}~\bibnamefont {Lupi}}, \bibinfo {author} {\bibfnamefont
  {P.~D.}\ \bibnamefont {Pietro}}, \bibinfo {author} {\bibfnamefont
  {D.}~\bibnamefont {Pontiroli}}, \bibinfo {author} {\bibfnamefont
  {M.}~\bibnamefont {Riccò}}, \bibinfo {author} {\bibfnamefont {S.~R.}\
  \bibnamefont {Clark}}, \bibinfo {author} {\bibfnamefont {D.}~\bibnamefont
  {Jaksch}}, \ and\ \bibinfo {author} {\bibfnamefont {A.}~\bibnamefont
  {Cavalleri}},\ }\href {\doibase 10.1038/nature16522} {\bibfield  {journal}
  {\bibinfo  {journal} {Nature}\ }\textbf {\bibinfo {volume} {530}},\ \bibinfo
  {pages} {461} (\bibinfo {year} {2016})}\BibitemShut {NoStop}%
\bibitem [{\citenamefont {Carusotto}\ and\ \citenamefont
  {Ciuti}(2013)}]{carusotto2013quantum}%
  \BibitemOpen
  \bibfield  {author} {\bibinfo {author} {\bibfnamefont {I.}~\bibnamefont
  {Carusotto}}\ and\ \bibinfo {author} {\bibfnamefont {C.}~\bibnamefont
  {Ciuti}},\ }\href {\doibase 10.1103/RevModPhys.85.299} {\bibfield  {journal}
  {\bibinfo  {journal} {Rev. Mod. Phys.}\ }\textbf {\bibinfo {volume} {85}},\
  \bibinfo {pages} {299} (\bibinfo {year} {2013})}\BibitemShut {NoStop}%
\bibitem [{\citenamefont {Schmidt}\ and\ \citenamefont
  {Koch}(2013)}]{schmidt2013circuit}%
  \BibitemOpen
  \bibfield  {author} {\bibinfo {author} {\bibfnamefont {S.}~\bibnamefont
  {Schmidt}}\ and\ \bibinfo {author} {\bibfnamefont {J.}~\bibnamefont {Koch}},\
  }\href {\doibase doi/10.1002/andp.201200261/full} {\bibfield  {journal}
  {\bibinfo  {journal} {Ann. Phys. (Berlin)}\ }\textbf {\bibinfo {volume}
  {525}},\ \bibinfo {pages} {395} (\bibinfo {year} {2013})}\BibitemShut
  {NoStop}%
\bibitem [{\citenamefont {Noh}\ and\ \citenamefont
  {Angelakis}(2016)}]{noh2016quantum}%
  \BibitemOpen
  \bibfield  {author} {\bibinfo {author} {\bibfnamefont {C.}~\bibnamefont
  {Noh}}\ and\ \bibinfo {author} {\bibfnamefont {D.~G.}\ \bibnamefont
  {Angelakis}},\ }\href {\doibase 10.1088/0034-4885/80/1/016401/meta}
  {\bibfield  {journal} {\bibinfo  {journal} {Rep. Prog. Phys.}\ }\textbf
  {\bibinfo {volume} {80}},\ \bibinfo {pages} {016401} (\bibinfo {year}
  {2016})}\BibitemShut {NoStop}%
\bibitem [{\citenamefont {Hartmann}(2016)}]{hartmann16}%
  \BibitemOpen
  \bibfield  {author} {\bibinfo {author} {\bibfnamefont {M.~J.}\ \bibnamefont
  {Hartmann}},\ }\href {\doibase 10.1088/2040-8978/18/10/104005} {\bibfield
  {journal} {\bibinfo  {journal} {Journal of Optics}\ }\textbf {\bibinfo
  {volume} {18}},\ \bibinfo {pages} {104005} (\bibinfo {year}
  {2016})}\BibitemShut {NoStop}%
\bibitem [{\citenamefont {Majumdar}\ \emph {et~al.}(2012)\citenamefont
  {Majumdar}, \citenamefont {Rundquist}, \citenamefont {Bajcsy},\ and\
  \citenamefont {Vu{\v{c}}kovi{\'c}}}]{majumdar2012cavity}%
  \BibitemOpen
  \bibfield  {author} {\bibinfo {author} {\bibfnamefont {A.}~\bibnamefont
  {Majumdar}}, \bibinfo {author} {\bibfnamefont {A.}~\bibnamefont {Rundquist}},
  \bibinfo {author} {\bibfnamefont {M.}~\bibnamefont {Bajcsy}}, \ and\ \bibinfo
  {author} {\bibfnamefont {J.}~\bibnamefont {Vu{\v{c}}kovi{\'c}}},\ }\href
  {\doibase /10.1103/PhysRevB.86.045315} {\bibfield  {journal} {\bibinfo
  {journal} {Phys. Rev. B}\ }\textbf {\bibinfo {volume} {86}},\ \bibinfo
  {pages} {045315} (\bibinfo {year} {2012})}\BibitemShut {NoStop}%
\bibitem [{\citenamefont {Sala}\ \emph {et~al.}(2015)\citenamefont {Sala},
  \citenamefont {Solnyshkov}, \citenamefont {Carusotto}, \citenamefont
  {Jacqmin}, \citenamefont {Lema\^{\i}tre}, \citenamefont
  {Ter\ifmmode~\mbox{\c{c}}\else \c{c}\fi{}as}, \citenamefont {Nalitov},
  \citenamefont {Abbarchi}, \citenamefont {Galopin}, \citenamefont {Sagnes},
  \citenamefont {Bloch}, \citenamefont {Malpuech},\ and\ \citenamefont
  {Amo}}]{sala2015vg}%
  \BibitemOpen
  \bibfield  {author} {\bibinfo {author} {\bibfnamefont {V.~G.}\ \bibnamefont
  {Sala}}, \bibinfo {author} {\bibfnamefont {D.~D.}\ \bibnamefont
  {Solnyshkov}}, \bibinfo {author} {\bibfnamefont {I.}~\bibnamefont
  {Carusotto}}, \bibinfo {author} {\bibfnamefont {T.}~\bibnamefont {Jacqmin}},
  \bibinfo {author} {\bibfnamefont {A.}~\bibnamefont {Lema\^{\i}tre}}, \bibinfo
  {author} {\bibfnamefont {H.}~\bibnamefont {Ter\ifmmode~\mbox{\c{c}}\else
  \c{c}\fi{}as}}, \bibinfo {author} {\bibfnamefont {A.}~\bibnamefont
  {Nalitov}}, \bibinfo {author} {\bibfnamefont {M.}~\bibnamefont {Abbarchi}},
  \bibinfo {author} {\bibfnamefont {E.}~\bibnamefont {Galopin}}, \bibinfo
  {author} {\bibfnamefont {I.}~\bibnamefont {Sagnes}}, \bibinfo {author}
  {\bibfnamefont {J.}~\bibnamefont {Bloch}}, \bibinfo {author} {\bibfnamefont
  {G.}~\bibnamefont {Malpuech}}, \ and\ \bibinfo {author} {\bibfnamefont
  {A.}~\bibnamefont {Amo}},\ }\href {\doibase 10.1103/PhysRevX.5.011034}
  {\bibfield  {journal} {\bibinfo  {journal} {Phys. Rev. X}\ }\textbf {\bibinfo
  {volume} {5}},\ \bibinfo {pages} {011034} (\bibinfo {year}
  {2015})}\BibitemShut {NoStop}%
\bibitem [{\citenamefont {Toyoda}\ \emph {et~al.}(2013)\citenamefont {Toyoda},
  \citenamefont {Matsuno}, \citenamefont {Noguchi}, \citenamefont {Haze},\ and\
  \citenamefont {Urabe}}]{toyoda2013experimental}%
  \BibitemOpen
  \bibfield  {author} {\bibinfo {author} {\bibfnamefont {K.}~\bibnamefont
  {Toyoda}}, \bibinfo {author} {\bibfnamefont {Y.}~\bibnamefont {Matsuno}},
  \bibinfo {author} {\bibfnamefont {A.}~\bibnamefont {Noguchi}}, \bibinfo
  {author} {\bibfnamefont {S.}~\bibnamefont {Haze}}, \ and\ \bibinfo {author}
  {\bibfnamefont {S.}~\bibnamefont {Urabe}},\ }\href {\doibase
  10.1103/PhysRevLett.111.160501} {\bibfield  {journal} {\bibinfo  {journal}
  {Phys. Rev. Lett.}\ }\textbf {\bibinfo {volume} {111}},\ \bibinfo {pages}
  {160501} (\bibinfo {year} {2013})}\BibitemShut {NoStop}%
\bibitem [{\citenamefont {Houck}\ \emph {et~al.}(2012)\citenamefont {Houck},
  \citenamefont {T{\"u}reci},\ and\ \citenamefont {Koch}}]{houck2012chip}%
  \BibitemOpen
  \bibfield  {author} {\bibinfo {author} {\bibfnamefont {A.~A.}\ \bibnamefont
  {Houck}}, \bibinfo {author} {\bibfnamefont {H.~E.}\ \bibnamefont
  {T{\"u}reci}}, \ and\ \bibinfo {author} {\bibfnamefont {J.}~\bibnamefont
  {Koch}},\ }\href {\doibase 10.1038/nphys2251} {\bibfield  {journal} {\bibinfo
   {journal} {Nature Physics}\ }\textbf {\bibinfo {volume} {8}},\ \bibinfo
  {pages} {292} (\bibinfo {year} {2012})}\BibitemShut {NoStop}%
\bibitem [{\citenamefont {Fitzpatrick}\ \emph {et~al.}(2017)\citenamefont
  {Fitzpatrick}, \citenamefont {Sundaresan}, \citenamefont {Li}, \citenamefont
  {Koch},\ and\ \citenamefont {Houck}}]{fitzpatrick17}%
  \BibitemOpen
  \bibfield  {author} {\bibinfo {author} {\bibfnamefont {M.}~\bibnamefont
  {Fitzpatrick}}, \bibinfo {author} {\bibfnamefont {N.~M.}\ \bibnamefont
  {Sundaresan}}, \bibinfo {author} {\bibfnamefont {A.~C.~Y.}\ \bibnamefont
  {Li}}, \bibinfo {author} {\bibfnamefont {J.}~\bibnamefont {Koch}}, \ and\
  \bibinfo {author} {\bibfnamefont {A.~A.}\ \bibnamefont {Houck}},\ }\href
  {\doibase 10.1103/PhysRevX.7.011016} {\bibfield  {journal} {\bibinfo
  {journal} {Phys. Rev. X}\ }\textbf {\bibinfo {volume} {7}},\ \bibinfo {pages}
  {011016} (\bibinfo {year} {2017})}\BibitemShut {NoStop}%
\bibitem [{\citenamefont {Carusotto}\ \emph {et~al.}(2009)\citenamefont
  {Carusotto}, \citenamefont {Gerace}, \citenamefont {Tureci}, \citenamefont
  {De~Liberato}, \citenamefont {Ciuti},\ and\ \citenamefont
  {{\.I}mamo{\u{g}}lu}}]{carusotto2009fermionized}%
  \BibitemOpen
  \bibfield  {author} {\bibinfo {author} {\bibfnamefont {I.}~\bibnamefont
  {Carusotto}}, \bibinfo {author} {\bibfnamefont {D.}~\bibnamefont {Gerace}},
  \bibinfo {author} {\bibfnamefont {H.}~\bibnamefont {Tureci}}, \bibinfo
  {author} {\bibfnamefont {S.}~\bibnamefont {De~Liberato}}, \bibinfo {author}
  {\bibfnamefont {C.}~\bibnamefont {Ciuti}}, \ and\ \bibinfo {author}
  {\bibfnamefont {A.}~\bibnamefont {{\.I}mamo{\u{g}}lu}},\ }\href {\doibase
  10.1103/PhysRevLett.103.033601} {\bibfield  {journal} {\bibinfo  {journal}
  {Phys. Rev. Lett.}\ }\textbf {\bibinfo {volume} {103}},\ \bibinfo {pages}
  {033601} (\bibinfo {year} {2009})}\BibitemShut {NoStop}%
\bibitem [{\citenamefont {Hartmann}(2010)}]{hartmann2010polariton}%
  \BibitemOpen
  \bibfield  {author} {\bibinfo {author} {\bibfnamefont {M.~J.}\ \bibnamefont
  {Hartmann}},\ }\href {\doibase 10.1103/PhysRevLett.104.113601} {\bibfield
  {journal} {\bibinfo  {journal} {Phys. Rev. Lett.}\ }\textbf {\bibinfo
  {volume} {104}},\ \bibinfo {pages} {113601} (\bibinfo {year}
  {2010})}\BibitemShut {NoStop}%
\bibitem [{\citenamefont {Grujic}\ \emph {et~al.}(2012)\citenamefont {Grujic},
  \citenamefont {Clark}, \citenamefont {Jaksch},\ and\ \citenamefont
  {Angelakis}}]{grujic2012non}%
  \BibitemOpen
  \bibfield  {author} {\bibinfo {author} {\bibfnamefont {T.}~\bibnamefont
  {Grujic}}, \bibinfo {author} {\bibfnamefont {S.}~\bibnamefont {Clark}},
  \bibinfo {author} {\bibfnamefont {D.}~\bibnamefont {Jaksch}}, \ and\ \bibinfo
  {author} {\bibfnamefont {D.}~\bibnamefont {Angelakis}},\ }\href {\doibase
  10.1088/1367-2630/14/10/103025/meta} {\bibfield  {journal} {\bibinfo
  {journal} {New J. Phys.}\ }\textbf {\bibinfo {volume} {14}},\ \bibinfo
  {pages} {103025} (\bibinfo {year} {2012})}\BibitemShut {NoStop}%
\bibitem [{\citenamefont {Nissen}\ \emph {et~al.}(2012)\citenamefont {Nissen},
  \citenamefont {Schmidt}, \citenamefont {Biondi}, \citenamefont {Blatter},
  \citenamefont {T{\"u}reci},\ and\ \citenamefont
  {Keeling}}]{nissen2012nonequilibrium}%
  \BibitemOpen
  \bibfield  {author} {\bibinfo {author} {\bibfnamefont {F.}~\bibnamefont
  {Nissen}}, \bibinfo {author} {\bibfnamefont {S.}~\bibnamefont {Schmidt}},
  \bibinfo {author} {\bibfnamefont {M.}~\bibnamefont {Biondi}}, \bibinfo
  {author} {\bibfnamefont {G.}~\bibnamefont {Blatter}}, \bibinfo {author}
  {\bibfnamefont {H.~E.}\ \bibnamefont {T{\"u}reci}}, \ and\ \bibinfo {author}
  {\bibfnamefont {J.}~\bibnamefont {Keeling}},\ }\href {\doibase
  10.1103/PhysRevLett.108.233603} {\bibfield  {journal} {\bibinfo  {journal}
  {Phys. Rev. Lett.}\ }\textbf {\bibinfo {volume} {108}},\ \bibinfo {pages}
  {233603} (\bibinfo {year} {2012})}\BibitemShut {NoStop}%
\bibitem [{\citenamefont {Joshi}\ \emph {et~al.}(2013)\citenamefont {Joshi},
  \citenamefont {Nissen},\ and\ \citenamefont {Keeling}}]{joshi2013quantum}%
  \BibitemOpen
  \bibfield  {author} {\bibinfo {author} {\bibfnamefont {C.}~\bibnamefont
  {Joshi}}, \bibinfo {author} {\bibfnamefont {F.}~\bibnamefont {Nissen}}, \
  and\ \bibinfo {author} {\bibfnamefont {J.}~\bibnamefont {Keeling}},\ }\href
  {\doibase 10.1103/PhysRevA.88.063835} {\bibfield  {journal} {\bibinfo
  {journal} {Phys. Rev.A}\ }\textbf {\bibinfo {volume} {88}},\ \bibinfo {pages}
  {063835} (\bibinfo {year} {2013})}\BibitemShut {NoStop}%
\bibitem [{\citenamefont {Jin}\ \emph {et~al.}(2013)\citenamefont {Jin},
  \citenamefont {Rossini}, \citenamefont {Fazio}, \citenamefont {Leib},\ and\
  \citenamefont {Hartmann}}]{jin2013photon}%
  \BibitemOpen
  \bibfield  {author} {\bibinfo {author} {\bibfnamefont {J.}~\bibnamefont
  {Jin}}, \bibinfo {author} {\bibfnamefont {D.}~\bibnamefont {Rossini}},
  \bibinfo {author} {\bibfnamefont {R.}~\bibnamefont {Fazio}}, \bibinfo
  {author} {\bibfnamefont {M.}~\bibnamefont {Leib}}, \ and\ \bibinfo {author}
  {\bibfnamefont {M.~J.}\ \bibnamefont {Hartmann}},\ }\href {\doibase
  10.1103/PhysRevLett.110.163605} {\bibfield  {journal} {\bibinfo  {journal}
  {Phys. Rev. Lett.}\ }\textbf {\bibinfo {volume} {110}},\ \bibinfo {pages}
  {163605} (\bibinfo {year} {2013})}\BibitemShut {NoStop}%
\bibitem [{\citenamefont {Jin}\ \emph {et~al.}(2014)\citenamefont {Jin},
  \citenamefont {Rossini}, \citenamefont {Leib}, \citenamefont {Hartmann},\
  and\ \citenamefont {Fazio}}]{jin2014steady}%
  \BibitemOpen
  \bibfield  {author} {\bibinfo {author} {\bibfnamefont {J.}~\bibnamefont
  {Jin}}, \bibinfo {author} {\bibfnamefont {D.}~\bibnamefont {Rossini}},
  \bibinfo {author} {\bibfnamefont {M.}~\bibnamefont {Leib}}, \bibinfo {author}
  {\bibfnamefont {M.~J.}\ \bibnamefont {Hartmann}}, \ and\ \bibinfo {author}
  {\bibfnamefont {R.}~\bibnamefont {Fazio}},\ }\href {\doibase
  10.1103/PhysRevA.90.023827} {\bibfield  {journal} {\bibinfo  {journal} {Phys.
  Rev.A}\ }\textbf {\bibinfo {volume} {90}},\ \bibinfo {pages} {023827}
  (\bibinfo {year} {2014})}\BibitemShut {NoStop}%
\bibitem [{\citenamefont {Biella}\ \emph {et~al.}(2015)\citenamefont {Biella},
  \citenamefont {Mazza}, \citenamefont {Carusotto}, \citenamefont {Rossini},\
  and\ \citenamefont {Fazio}}]{biella2015photon}%
  \BibitemOpen
  \bibfield  {author} {\bibinfo {author} {\bibfnamefont {A.}~\bibnamefont
  {Biella}}, \bibinfo {author} {\bibfnamefont {L.}~\bibnamefont {Mazza}},
  \bibinfo {author} {\bibfnamefont {I.}~\bibnamefont {Carusotto}}, \bibinfo
  {author} {\bibfnamefont {D.}~\bibnamefont {Rossini}}, \ and\ \bibinfo
  {author} {\bibfnamefont {R.}~\bibnamefont {Fazio}},\ }\href {\doibase
  10.1103/PhysRevA.91.053815} {\bibfield  {journal} {\bibinfo  {journal} {Phys.
  Rev.A}\ }\textbf {\bibinfo {volume} {91}},\ \bibinfo {pages} {053815}
  (\bibinfo {year} {2015})}\BibitemShut {NoStop}%
\bibitem [{\citenamefont {Schir{\'o}}\ \emph {et~al.}(2016)\citenamefont
  {Schir{\'o}}, \citenamefont {Joshi}, \citenamefont {Bordyuh}, \citenamefont
  {Fazio}, \citenamefont {Keeling},\ and\ \citenamefont
  {T{\"u}reci}}]{schiro2016exotic}%
  \BibitemOpen
  \bibfield  {author} {\bibinfo {author} {\bibfnamefont {M.}~\bibnamefont
  {Schir{\'o}}}, \bibinfo {author} {\bibfnamefont {C.}~\bibnamefont {Joshi}},
  \bibinfo {author} {\bibfnamefont {M.}~\bibnamefont {Bordyuh}}, \bibinfo
  {author} {\bibfnamefont {R.}~\bibnamefont {Fazio}}, \bibinfo {author}
  {\bibfnamefont {J.}~\bibnamefont {Keeling}}, \ and\ \bibinfo {author}
  {\bibfnamefont {H.}~\bibnamefont {T{\"u}reci}},\ }\href {\doibase
  10.1103/PhysRevLett.116.143603} {\bibfield  {journal} {\bibinfo  {journal}
  {Phys. Rev. Lett.}\ }\textbf {\bibinfo {volume} {116}},\ \bibinfo {pages}
  {143603} (\bibinfo {year} {2016})}\BibitemShut {NoStop}%
\bibitem [{\citenamefont {Lee}\ \emph {et~al.}(2013)\citenamefont {Lee},
  \citenamefont {Gopalakrishnan},\ and\ \citenamefont
  {Lukin}}]{lee2013unconventional}%
  \BibitemOpen
  \bibfield  {author} {\bibinfo {author} {\bibfnamefont {T.~E.}\ \bibnamefont
  {Lee}}, \bibinfo {author} {\bibfnamefont {S.}~\bibnamefont {Gopalakrishnan}},
  \ and\ \bibinfo {author} {\bibfnamefont {M.~D.}\ \bibnamefont {Lukin}},\
  }\href {\doibase 10.1103/PhysRevLett.110.257204} {\bibfield  {journal}
  {\bibinfo  {journal} {Phys. Rev. Lett.}\ }\textbf {\bibinfo {volume} {110}},\
  \bibinfo {pages} {257204} (\bibinfo {year} {2013})}\BibitemShut {NoStop}%
\bibitem [{\citenamefont {Breuer}\ and\ \citenamefont
  {Petruccione}(2002)}]{breuer2002theory}%
  \BibitemOpen
  \bibfield  {author} {\bibinfo {author} {\bibfnamefont {H.-P.}\ \bibnamefont
  {Breuer}}\ and\ \bibinfo {author} {\bibfnamefont {F.}~\bibnamefont
  {Petruccione}},\ }\href {\doibase 10.1093/acprof:oso/9780199213900.001.0001}
  {\emph {\bibinfo {title} {The theory of open quantum systems}}}\ (\bibinfo
  {publisher} {Oxford University Press},\ \bibinfo {address} {Oxford},\
  \bibinfo {year} {2002})\BibitemShut {NoStop}%
\bibitem [{\citenamefont {Vamivakas}\ \emph {et~al.}(2009)\citenamefont
  {Vamivakas}, \citenamefont {Zhao}, \citenamefont {Lu},\ and\ \citenamefont
  {Atat{\"u}re}}]{vamivakas2009spin}%
  \BibitemOpen
  \bibfield  {author} {\bibinfo {author} {\bibfnamefont {A.~N.}\ \bibnamefont
  {Vamivakas}}, \bibinfo {author} {\bibfnamefont {Y.}~\bibnamefont {Zhao}},
  \bibinfo {author} {\bibfnamefont {C.-Y.}\ \bibnamefont {Lu}}, \ and\ \bibinfo
  {author} {\bibfnamefont {M.}~\bibnamefont {Atat{\"u}re}},\ }\href {\doibase
  10.1038/nphys1182} {\bibfield  {journal} {\bibinfo  {journal} {Nature
  Physics}\ }\textbf {\bibinfo {volume} {5}},\ \bibinfo {pages} {198} (\bibinfo
  {year} {2009})}\BibitemShut {NoStop}%
\bibitem [{\citenamefont {Lang}\ \emph {et~al.}(2011)\citenamefont {Lang},
  \citenamefont {Bozyigit}, \citenamefont {Eichler}, \citenamefont {Steffen},
  \citenamefont {Fink}, \citenamefont {Abdumalikov}, \citenamefont {Baur},
  \citenamefont {Filipp}, \citenamefont {da~Silva}, \citenamefont {Blais},\
  and\ \citenamefont {Wallraff}}]{lang11rf}%
  \BibitemOpen
  \bibfield  {author} {\bibinfo {author} {\bibfnamefont {C.}~\bibnamefont
  {Lang}}, \bibinfo {author} {\bibfnamefont {D.}~\bibnamefont {Bozyigit}},
  \bibinfo {author} {\bibfnamefont {C.}~\bibnamefont {Eichler}}, \bibinfo
  {author} {\bibfnamefont {L.}~\bibnamefont {Steffen}}, \bibinfo {author}
  {\bibfnamefont {J.~M.}\ \bibnamefont {Fink}}, \bibinfo {author}
  {\bibfnamefont {A.~A.}\ \bibnamefont {Abdumalikov}}, \bibinfo {author}
  {\bibfnamefont {M.}~\bibnamefont {Baur}}, \bibinfo {author} {\bibfnamefont
  {S.}~\bibnamefont {Filipp}}, \bibinfo {author} {\bibfnamefont {M.~P.}\
  \bibnamefont {da~Silva}}, \bibinfo {author} {\bibfnamefont {A.}~\bibnamefont
  {Blais}}, \ and\ \bibinfo {author} {\bibfnamefont {A.}~\bibnamefont
  {Wallraff}},\ }\href {\doibase 10.1103/PhysRevLett.106.243601} {\bibfield
  {journal} {\bibinfo  {journal} {Phys. Rev. Lett.}\ }\textbf {\bibinfo
  {volume} {106}},\ \bibinfo {pages} {243601} (\bibinfo {year}
  {2011})}\BibitemShut {NoStop}%
\bibitem [{\citenamefont {Liew}\ and\ \citenamefont
  {Savona}(2010)}]{liew10single}%
  \BibitemOpen
  \bibfield  {author} {\bibinfo {author} {\bibfnamefont {T.~C.~H.}\
  \bibnamefont {Liew}}\ and\ \bibinfo {author} {\bibfnamefont {V.}~\bibnamefont
  {Savona}},\ }\href {\doibase 10.1103/PhysRevLett.104.183601} {\bibfield
  {journal} {\bibinfo  {journal} {Phys. Rev. Lett.}\ }\textbf {\bibinfo
  {volume} {104}},\ \bibinfo {pages} {183601} (\bibinfo {year}
  {2010})}\BibitemShut {NoStop}%
\bibitem [{\citenamefont {Bamba}\ \emph {et~al.}(2011)\citenamefont {Bamba},
  \citenamefont {Imamo\ifmmode~\breve{g}\else \u{g}\fi{}lu}, \citenamefont
  {Carusotto},\ and\ \citenamefont {Ciuti}}]{bamba11single}%
  \BibitemOpen
  \bibfield  {author} {\bibinfo {author} {\bibfnamefont {M.}~\bibnamefont
  {Bamba}}, \bibinfo {author} {\bibfnamefont {A.}~\bibnamefont
  {Imamo\ifmmode~\breve{g}\else \u{g}\fi{}lu}}, \bibinfo {author}
  {\bibfnamefont {I.}~\bibnamefont {Carusotto}}, \ and\ \bibinfo {author}
  {\bibfnamefont {C.}~\bibnamefont {Ciuti}},\ }\href {\doibase
  10.1103/PhysRevA.83.021802} {\bibfield  {journal} {\bibinfo  {journal} {Phys.
  Rev. A}\ }\textbf {\bibinfo {volume} {83}},\ \bibinfo {pages} {021802}
  (\bibinfo {year} {2011})}\BibitemShut {NoStop}%
\bibitem [{\citenamefont {Rodriguez}\ \emph {et~al.}(2017)\citenamefont
  {Rodriguez}, \citenamefont {Casteels}, \citenamefont {Storme}, \citenamefont
  {Carlon~Zambon}, \citenamefont {Sagnes}, \citenamefont {Le~Gratiet},
  \citenamefont {Galopin}, \citenamefont {Lema\^{\i}tre}, \citenamefont {Amo},
  \citenamefont {Ciuti},\ and\ \citenamefont {Bloch}}]{rodriguez17hysteresis}%
  \BibitemOpen
  \bibfield  {author} {\bibinfo {author} {\bibfnamefont {S.~R.~K.}\
  \bibnamefont {Rodriguez}}, \bibinfo {author} {\bibfnamefont {W.}~\bibnamefont
  {Casteels}}, \bibinfo {author} {\bibfnamefont {F.}~\bibnamefont {Storme}},
  \bibinfo {author} {\bibfnamefont {N.}~\bibnamefont {Carlon~Zambon}}, \bibinfo
  {author} {\bibfnamefont {I.}~\bibnamefont {Sagnes}}, \bibinfo {author}
  {\bibfnamefont {L.}~\bibnamefont {Le~Gratiet}}, \bibinfo {author}
  {\bibfnamefont {E.}~\bibnamefont {Galopin}}, \bibinfo {author} {\bibfnamefont
  {A.}~\bibnamefont {Lema\^{\i}tre}}, \bibinfo {author} {\bibfnamefont
  {A.}~\bibnamefont {Amo}}, \bibinfo {author} {\bibfnamefont {C.}~\bibnamefont
  {Ciuti}}, \ and\ \bibinfo {author} {\bibfnamefont {J.}~\bibnamefont
  {Bloch}},\ }\href {\doibase 10.1103/PhysRevLett.118.247402} {\bibfield
  {journal} {\bibinfo  {journal} {Phys. Rev. Lett.}\ }\textbf {\bibinfo
  {volume} {118}},\ \bibinfo {pages} {247402} (\bibinfo {year}
  {2017})}\BibitemShut {NoStop}%
\bibitem [{\citenamefont {Fink}\ \emph {et~al.}(2018)\citenamefont {Fink},
  \citenamefont {Schade}, \citenamefont {H{\"o}fling}, \citenamefont
  {Schneider},\ and\ \citenamefont {{\.I}mamo{\u{g}}lu}}]{fink2017signatures}%
  \BibitemOpen
  \bibfield  {author} {\bibinfo {author} {\bibfnamefont {T.}~\bibnamefont
  {Fink}}, \bibinfo {author} {\bibfnamefont {A.}~\bibnamefont {Schade}},
  \bibinfo {author} {\bibfnamefont {S.}~\bibnamefont {H{\"o}fling}}, \bibinfo
  {author} {\bibfnamefont {C.}~\bibnamefont {Schneider}}, \ and\ \bibinfo
  {author} {\bibfnamefont {A.}~\bibnamefont {{\.I}mamo{\u{g}}lu}},\ }\href
  {\doibase 10.1038/s41567-017-0020-9} {\bibfield  {journal} {\bibinfo
  {journal} {Nature Physics}\ }\textbf {\bibinfo {volume} {14}},\ \bibinfo
  {pages} {365} (\bibinfo {year} {2018})}\BibitemShut {NoStop}%
\bibitem [{\citenamefont {Chan}\ \emph {et~al.}(2015)\citenamefont {Chan},
  \citenamefont {Lee},\ and\ \citenamefont {Gopalakrishnan}}]{chan2015limit}%
  \BibitemOpen
  \bibfield  {author} {\bibinfo {author} {\bibfnamefont {C.-K.}\ \bibnamefont
  {Chan}}, \bibinfo {author} {\bibfnamefont {T.~E.}\ \bibnamefont {Lee}}, \
  and\ \bibinfo {author} {\bibfnamefont {S.}~\bibnamefont {Gopalakrishnan}},\
  }\href {\doibase 10.1103/PhysRevA.91.051601} {\bibfield  {journal} {\bibinfo
  {journal} {Phys. Rev.A}\ }\textbf {\bibinfo {volume} {91}},\ \bibinfo {pages}
  {051601} (\bibinfo {year} {2015})}\BibitemShut {NoStop}%
\bibitem [{\citenamefont {Iemini}\ \emph {et~al.}(2018)\citenamefont {Iemini},
  \citenamefont {Russomanno}, \citenamefont {Keeling}, \citenamefont
  {Schir\`o}, \citenamefont {Dalmonte},\ and\ \citenamefont
  {Fazio}}]{iemini2017boundary}%
  \BibitemOpen
  \bibfield  {author} {\bibinfo {author} {\bibfnamefont {F.}~\bibnamefont
  {Iemini}}, \bibinfo {author} {\bibfnamefont {A.}~\bibnamefont {Russomanno}},
  \bibinfo {author} {\bibfnamefont {J.}~\bibnamefont {Keeling}}, \bibinfo
  {author} {\bibfnamefont {M.}~\bibnamefont {Schir\`o}}, \bibinfo {author}
  {\bibfnamefont {M.}~\bibnamefont {Dalmonte}}, \ and\ \bibinfo {author}
  {\bibfnamefont {R.}~\bibnamefont {Fazio}},\ }\href {\doibase
  10.1103/PhysRevLett.121.035301} {\bibfield  {journal} {\bibinfo  {journal}
  {Phys. Rev. Lett.}\ }\textbf {\bibinfo {volume} {121}},\ \bibinfo {pages}
  {035301} (\bibinfo {year} {2018})}\BibitemShut {NoStop}%
\bibitem [{\citenamefont {Diehl}\ \emph {et~al.}(2008)\citenamefont {Diehl},
  \citenamefont {Micheli}, \citenamefont {Kantian}, \citenamefont {Kraus},
  \citenamefont {B{\"u}chler},\ and\ \citenamefont
  {Zoller}}]{diehl2008quantum}%
  \BibitemOpen
  \bibfield  {author} {\bibinfo {author} {\bibfnamefont {S.}~\bibnamefont
  {Diehl}}, \bibinfo {author} {\bibfnamefont {A.}~\bibnamefont {Micheli}},
  \bibinfo {author} {\bibfnamefont {A.}~\bibnamefont {Kantian}}, \bibinfo
  {author} {\bibfnamefont {B.}~\bibnamefont {Kraus}}, \bibinfo {author}
  {\bibfnamefont {H.}~\bibnamefont {B{\"u}chler}}, \ and\ \bibinfo {author}
  {\bibfnamefont {P.}~\bibnamefont {Zoller}},\ }\href {\doibase
  doi:10.1038/nphys1073} {\bibfield  {journal} {\bibinfo  {journal} {Nat.
  Phys.}\ }\textbf {\bibinfo {volume} {4}},\ \bibinfo {pages} {878} (\bibinfo
  {year} {2008})}\BibitemShut {NoStop}%
\bibitem [{\citenamefont {Diehl}\ \emph {et~al.}(2010)\citenamefont {Diehl},
  \citenamefont {Tomadin}, \citenamefont {Micheli}, \citenamefont {Fazio},\
  and\ \citenamefont {Zoller}}]{diehl2010dynamical}%
  \BibitemOpen
  \bibfield  {author} {\bibinfo {author} {\bibfnamefont {S.}~\bibnamefont
  {Diehl}}, \bibinfo {author} {\bibfnamefont {A.}~\bibnamefont {Tomadin}},
  \bibinfo {author} {\bibfnamefont {A.}~\bibnamefont {Micheli}}, \bibinfo
  {author} {\bibfnamefont {R.}~\bibnamefont {Fazio}}, \ and\ \bibinfo {author}
  {\bibfnamefont {P.}~\bibnamefont {Zoller}},\ }\href {\doibase
  10.1103/PhysRevLett.105.015702} {\bibfield  {journal} {\bibinfo  {journal}
  {Phys. Rev. Lett.}\ }\textbf {\bibinfo {volume} {105}},\ \bibinfo {pages}
  {015702} (\bibinfo {year} {2010})}\BibitemShut {NoStop}%
\bibitem [{\citenamefont {{\"O}ztop}\ \emph {et~al.}(2012)\citenamefont
  {{\"O}ztop}, \citenamefont {Bordyuh}, \citenamefont
  {M{\"u}stecapl{\i}o{\u{g}}lu},\ and\ \citenamefont
  {T{\"u}reci}}]{oztop2012excitations}%
  \BibitemOpen
  \bibfield  {author} {\bibinfo {author} {\bibfnamefont {B.}~\bibnamefont
  {{\"O}ztop}}, \bibinfo {author} {\bibfnamefont {M.}~\bibnamefont {Bordyuh}},
  \bibinfo {author} {\bibfnamefont {{\"O}.~E.}\ \bibnamefont
  {M{\"u}stecapl{\i}o{\u{g}}lu}}, \ and\ \bibinfo {author} {\bibfnamefont
  {H.~E.}\ \bibnamefont {T{\"u}reci}},\ }\href {\doibase
  10.1088/1367-2630/14/8/085011/meta} {\bibfield  {journal} {\bibinfo
  {journal} {New J. Phys.}\ }\textbf {\bibinfo {volume} {14}},\ \bibinfo
  {pages} {085011} (\bibinfo {year} {2012})}\BibitemShut {NoStop}%
\bibitem [{\citenamefont {Torre}\ \emph {et~al.}(2013)\citenamefont {Torre},
  \citenamefont {Diehl}, \citenamefont {Lukin}, \citenamefont {Sachdev},\ and\
  \citenamefont {Strack}}]{dalla2013keldysh}%
  \BibitemOpen
  \bibfield  {author} {\bibinfo {author} {\bibfnamefont {E.~G.~D.}\
  \bibnamefont {Torre}}, \bibinfo {author} {\bibfnamefont {S.}~\bibnamefont
  {Diehl}}, \bibinfo {author} {\bibfnamefont {M.~D.}\ \bibnamefont {Lukin}},
  \bibinfo {author} {\bibfnamefont {S.}~\bibnamefont {Sachdev}}, \ and\
  \bibinfo {author} {\bibfnamefont {P.}~\bibnamefont {Strack}},\ }\href
  {\doibase 10.1103/PhysRevA.87.023831} {\bibfield  {journal} {\bibinfo
  {journal} {Phys. Rev.A}\ }\textbf {\bibinfo {volume} {87}},\ \bibinfo {pages}
  {023831} (\bibinfo {year} {2013})}\BibitemShut {NoStop}%
\bibitem [{\citenamefont {Buchhold}\ \emph {et~al.}(2013)\citenamefont
  {Buchhold}, \citenamefont {Strack}, \citenamefont {Sachdev},\ and\
  \citenamefont {Diehl}}]{buchhold2013dicke}%
  \BibitemOpen
  \bibfield  {author} {\bibinfo {author} {\bibfnamefont {M.}~\bibnamefont
  {Buchhold}}, \bibinfo {author} {\bibfnamefont {P.}~\bibnamefont {Strack}},
  \bibinfo {author} {\bibfnamefont {S.}~\bibnamefont {Sachdev}}, \ and\
  \bibinfo {author} {\bibfnamefont {S.}~\bibnamefont {Diehl}},\ }\href
  {\doibase 10.1103/PhysRevA.87.063622} {\bibfield  {journal} {\bibinfo
  {journal} {Phys. Rev.A}\ }\textbf {\bibinfo {volume} {87}},\ \bibinfo {pages}
  {063622} (\bibinfo {year} {2013})}\BibitemShut {NoStop}%
\bibitem [{\citenamefont {Sieberer}\ \emph {et~al.}(2013)\citenamefont
  {Sieberer}, \citenamefont {Huber}, \citenamefont {Altman},\ and\
  \citenamefont {Diehl}}]{sieberer2013dynamical}%
  \BibitemOpen
  \bibfield  {author} {\bibinfo {author} {\bibfnamefont {L.}~\bibnamefont
  {Sieberer}}, \bibinfo {author} {\bibfnamefont {S.}~\bibnamefont {Huber}},
  \bibinfo {author} {\bibfnamefont {E.}~\bibnamefont {Altman}}, \ and\ \bibinfo
  {author} {\bibfnamefont {S.}~\bibnamefont {Diehl}},\ }\href {\doibase
  10.1103/PhysRevLett.110.195301} {\bibfield  {journal} {\bibinfo  {journal}
  {Phys. Rev. Lett.}\ }\textbf {\bibinfo {volume} {110}},\ \bibinfo {pages}
  {195301} (\bibinfo {year} {2013})}\BibitemShut {NoStop}%
\bibitem [{\citenamefont {Sieberer}\ \emph {et~al.}(2016)\citenamefont
  {Sieberer}, \citenamefont {Buchhold},\ and\ \citenamefont
  {Diehl}}]{sieberer2016keldysh}%
  \BibitemOpen
  \bibfield  {author} {\bibinfo {author} {\bibfnamefont {L.~M.}\ \bibnamefont
  {Sieberer}}, \bibinfo {author} {\bibfnamefont {M.}~\bibnamefont {Buchhold}},
  \ and\ \bibinfo {author} {\bibfnamefont {S.}~\bibnamefont {Diehl}},\ }\href
  {\doibase 10.1088/0034-4885/79/9/096001/meta} {\bibfield  {journal} {\bibinfo
   {journal} {Rep. Prog. Phys.}\ }\textbf {\bibinfo {volume} {79}},\ \bibinfo
  {pages} {096001} (\bibinfo {year} {2016})}\BibitemShut {NoStop}%
\bibitem [{\citenamefont {Klaers}\ \emph {et~al.}(2010)\citenamefont {Klaers},
  \citenamefont {Schmitt}, \citenamefont {Vewinger},\ and\ \citenamefont
  {Weitz}}]{klaers2010bose}%
  \BibitemOpen
  \bibfield  {author} {\bibinfo {author} {\bibfnamefont {J.}~\bibnamefont
  {Klaers}}, \bibinfo {author} {\bibfnamefont {J.}~\bibnamefont {Schmitt}},
  \bibinfo {author} {\bibfnamefont {F.}~\bibnamefont {Vewinger}}, \ and\
  \bibinfo {author} {\bibfnamefont {M.}~\bibnamefont {Weitz}},\ }\href
  {\doibase doi:10.1038/nature09567} {\bibfield  {journal} {\bibinfo  {journal}
  {Nature}\ }\textbf {\bibinfo {volume} {468}},\ \bibinfo {pages} {545}
  (\bibinfo {year} {2010})}\BibitemShut {NoStop}%
\bibitem [{\citenamefont {Kirton}\ and\ \citenamefont
  {Keeling}(2015)}]{kirton2015thermalization}%
  \BibitemOpen
  \bibfield  {author} {\bibinfo {author} {\bibfnamefont {P.}~\bibnamefont
  {Kirton}}\ and\ \bibinfo {author} {\bibfnamefont {J.}~\bibnamefont
  {Keeling}},\ }\href {\doibase 10.1103/PhysRevA.91.033826} {\bibfield
  {journal} {\bibinfo  {journal} {Phys. Rev.A}\ }\textbf {\bibinfo {volume}
  {91}},\ \bibinfo {pages} {033826} (\bibinfo {year} {2015})}\BibitemShut
  {NoStop}%
\bibitem [{\citenamefont {Doan}\ \emph {et~al.}(2005)\citenamefont {Doan},
  \citenamefont {Cao}, \citenamefont {Tran~Thoai},\ and\ \citenamefont
  {Haug}}]{doan05}%
  \BibitemOpen
  \bibfield  {author} {\bibinfo {author} {\bibfnamefont {T.~D.}\ \bibnamefont
  {Doan}}, \bibinfo {author} {\bibfnamefont {H.~T.}\ \bibnamefont {Cao}},
  \bibinfo {author} {\bibfnamefont {D.~B.}\ \bibnamefont {Tran~Thoai}}, \ and\
  \bibinfo {author} {\bibfnamefont {H.}~\bibnamefont {Haug}},\ }\href {\doibase
  10.1103/PhysRevB.72.085301} {\bibfield  {journal} {\bibinfo  {journal} {Phys.
  Rev. B}\ }\textbf {\bibinfo {volume} {72}},\ \bibinfo {pages} {085301}
  (\bibinfo {year} {2005})}\BibitemShut {NoStop}%
\bibitem [{\citenamefont {Kasprzak}\ \emph {et~al.}(2008)\citenamefont
  {Kasprzak}, \citenamefont {Solnyshkov}, \citenamefont {Andr\'e},
  \citenamefont {Dang},\ and\ \citenamefont {Malpuech}}]{kasprzak08}%
  \BibitemOpen
  \bibfield  {author} {\bibinfo {author} {\bibfnamefont {J.}~\bibnamefont
  {Kasprzak}}, \bibinfo {author} {\bibfnamefont {D.~D.}\ \bibnamefont
  {Solnyshkov}}, \bibinfo {author} {\bibfnamefont {R.}~\bibnamefont {Andr\'e}},
  \bibinfo {author} {\bibfnamefont {L.~S.}\ \bibnamefont {Dang}}, \ and\
  \bibinfo {author} {\bibfnamefont {G.}~\bibnamefont {Malpuech}},\ }\href
  {\doibase 10.1103/PhysRevLett.101.146404} {\bibfield  {journal} {\bibinfo
  {journal} {Phys. Rev. Lett.}\ }\textbf {\bibinfo {volume} {101}},\ \bibinfo
  {pages} {146404} (\bibinfo {year} {2008})}\BibitemShut {NoStop}%
\bibitem [{\citenamefont {Sun}\ \emph {et~al.}(2017)\citenamefont {Sun},
  \citenamefont {Wen}, \citenamefont {Yoon}, \citenamefont {Liu}, \citenamefont
  {Steger}, \citenamefont {Pfeiffer}, \citenamefont {West}, \citenamefont
  {Snoke},\ and\ \citenamefont {Nelson}}]{sun17thermalisation}%
  \BibitemOpen
  \bibfield  {author} {\bibinfo {author} {\bibfnamefont {Y.}~\bibnamefont
  {Sun}}, \bibinfo {author} {\bibfnamefont {P.}~\bibnamefont {Wen}}, \bibinfo
  {author} {\bibfnamefont {Y.}~\bibnamefont {Yoon}}, \bibinfo {author}
  {\bibfnamefont {G.}~\bibnamefont {Liu}}, \bibinfo {author} {\bibfnamefont
  {M.}~\bibnamefont {Steger}}, \bibinfo {author} {\bibfnamefont {L.~N.}\
  \bibnamefont {Pfeiffer}}, \bibinfo {author} {\bibfnamefont {K.}~\bibnamefont
  {West}}, \bibinfo {author} {\bibfnamefont {D.~W.}\ \bibnamefont {Snoke}}, \
  and\ \bibinfo {author} {\bibfnamefont {K.~A.}\ \bibnamefont {Nelson}},\
  }\href {\doibase 10.1103/PhysRevLett.118.016602} {\bibfield  {journal}
  {\bibinfo  {journal} {Phys. Rev. Lett.}\ }\textbf {\bibinfo {volume} {118}},\
  \bibinfo {pages} {016602} (\bibinfo {year} {2017})}\BibitemShut {NoStop}%
\bibitem [{\citenamefont {Polkovnikov}\ \emph {et~al.}(2011)\citenamefont
  {Polkovnikov}, \citenamefont {Sengupta}, \citenamefont {Silva},\ and\
  \citenamefont {Vengalattore}}]{polkovnikov2011colloquium}%
  \BibitemOpen
  \bibfield  {author} {\bibinfo {author} {\bibfnamefont {A.}~\bibnamefont
  {Polkovnikov}}, \bibinfo {author} {\bibfnamefont {K.}~\bibnamefont
  {Sengupta}}, \bibinfo {author} {\bibfnamefont {A.}~\bibnamefont {Silva}}, \
  and\ \bibinfo {author} {\bibfnamefont {M.}~\bibnamefont {Vengalattore}},\
  }\href {\doibase 10.1103/RevModPhys.83.863} {\bibfield  {journal} {\bibinfo
  {journal} {Rev. Mod. Phys.}\ }\textbf {\bibinfo {volume} {83}},\ \bibinfo
  {pages} {863} (\bibinfo {year} {2011})}\BibitemShut {NoStop}%
\bibitem [{\citenamefont {Eisert}\ \emph {et~al.}(2015)\citenamefont {Eisert},
  \citenamefont {Friesdorf},\ and\ \citenamefont
  {Gogolin}}]{eisert2015quantum}%
  \BibitemOpen
  \bibfield  {author} {\bibinfo {author} {\bibfnamefont {J.}~\bibnamefont
  {Eisert}}, \bibinfo {author} {\bibfnamefont {M.}~\bibnamefont {Friesdorf}}, \
  and\ \bibinfo {author} {\bibfnamefont {C.}~\bibnamefont {Gogolin}},\ }\href
  {\doibase 10.1038/nphys3215} {\bibfield  {journal} {\bibinfo  {journal} {Nat.
  Phys.}\ }\textbf {\bibinfo {volume} {11}},\ \bibinfo {pages} {124} (\bibinfo
  {year} {2015})}\BibitemShut {NoStop}%
\bibitem [{\citenamefont {Langen}\ \emph {et~al.}(2016)\citenamefont {Langen},
  \citenamefont {Gasenzer},\ and\ \citenamefont
  {Schmiedmayer}}]{langen16pretherm}%
  \BibitemOpen
  \bibfield  {author} {\bibinfo {author} {\bibfnamefont {T.}~\bibnamefont
  {Langen}}, \bibinfo {author} {\bibfnamefont {T.}~\bibnamefont {Gasenzer}}, \
  and\ \bibinfo {author} {\bibfnamefont {J.}~\bibnamefont {Schmiedmayer}},\
  }\href@noop {} {\bibfield  {journal} {\bibinfo  {journal} {J. Stat. Mech.
  Theor. Exp.}\ }\textbf {\bibinfo {volume} {2016}},\ \bibinfo {pages} {064009}
  (\bibinfo {year} {2016})}\BibitemShut {NoStop}%
\bibitem [{\citenamefont {Wolff}\ \emph {et~al.}(2018)\citenamefont {Wolff},
  \citenamefont {Bernier}, \citenamefont {Poletti}, \citenamefont {Sheikhan},\
  and\ \citenamefont {Kollath}}]{wolff2018evolution}%
  \BibitemOpen
  \bibfield  {author} {\bibinfo {author} {\bibfnamefont {S.}~\bibnamefont
  {Wolff}}, \bibinfo {author} {\bibfnamefont {J.-S.}\ \bibnamefont {Bernier}},
  \bibinfo {author} {\bibfnamefont {D.}~\bibnamefont {Poletti}}, \bibinfo
  {author} {\bibfnamefont {A.}~\bibnamefont {Sheikhan}}, \ and\ \bibinfo
  {author} {\bibfnamefont {C.}~\bibnamefont {Kollath}},\ }\href@noop {}
  {\enquote {\bibinfo {title} {Evolution of two-time correlations in
  dissipative quantum spin systems: aging and hierarchical dynamics},}\ }
  (\bibinfo {year} {2018}),\ \bibinfo {note} {preprint},\ \Eprint
  {http://arxiv.org/abs/1809.10464} {1809.10464} \BibitemShut {NoStop}%
\bibitem [{\citenamefont {Bardyn}\ and\ \citenamefont
  {{\.I}mamo{\u{g}}lu}(2012)}]{bardyn2012majorana}%
  \BibitemOpen
  \bibfield  {author} {\bibinfo {author} {\bibfnamefont {C.-E.}\ \bibnamefont
  {Bardyn}}\ and\ \bibinfo {author} {\bibfnamefont {A.}~\bibnamefont
  {{\.I}mamo{\u{g}}lu}},\ }\href {\doibase 10.1103/PhysRevLett.109.253606}
  {\bibfield  {journal} {\bibinfo  {journal} {Phys. Rev. Lett.}\ }\textbf
  {\bibinfo {volume} {109}},\ \bibinfo {pages} {253606} (\bibinfo {year}
  {2012})}\BibitemShut {NoStop}%
\bibitem [{\citenamefont {Mascarenhas}\ \emph {et~al.}(2015)\citenamefont
  {Mascarenhas}, \citenamefont {Flayac},\ and\ \citenamefont
  {Savona}}]{mascarenhas2015matrix}%
  \BibitemOpen
  \bibfield  {author} {\bibinfo {author} {\bibfnamefont {E.}~\bibnamefont
  {Mascarenhas}}, \bibinfo {author} {\bibfnamefont {H.}~\bibnamefont {Flayac}},
  \ and\ \bibinfo {author} {\bibfnamefont {V.}~\bibnamefont {Savona}},\ }\href
  {\doibase 10.1103/PhysRevA.92.022116} {\bibfield  {journal} {\bibinfo
  {journal} {Phys. Rev.A}\ }\textbf {\bibinfo {volume} {92}},\ \bibinfo {pages}
  {022116} (\bibinfo {year} {2015})}\BibitemShut {NoStop}%
\bibitem [{sup()}]{supplementary}%
  \BibitemOpen
  \href@noop {} {}\bibinfo {note} {Supplementary Material containing derivation
  of the effective model, details of the spin wave calculation, and a full
  description of the tensor network method for two-time
  correlations.}\BibitemShut {Stop}%
\bibitem [{\citenamefont {Schollw{\"o}ck}(2011)}]{schollwock2011density}%
  \BibitemOpen
  \bibfield  {author} {\bibinfo {author} {\bibfnamefont {U.}~\bibnamefont
  {Schollw{\"o}ck}},\ }\href {\doibase /10.1016/j.aop.2010.09.012} {\bibfield
  {journal} {\bibinfo  {journal} {Ann. Phys.}\ }\textbf {\bibinfo {volume}
  {326}},\ \bibinfo {pages} {96} (\bibinfo {year} {2011})}\BibitemShut
  {NoStop}%
\bibitem [{\citenamefont {Or{\'u}s}(2014)}]{orus2014practical}%
  \BibitemOpen
  \bibfield  {author} {\bibinfo {author} {\bibfnamefont {R.}~\bibnamefont
  {Or{\'u}s}},\ }\href {\doibase 10.1016/j.aop.2014.06.013} {\bibfield
  {journal} {\bibinfo  {journal} {Ann. Phys.}\ }\textbf {\bibinfo {volume}
  {349}},\ \bibinfo {pages} {117} (\bibinfo {year} {2014})}\BibitemShut
  {NoStop}%
\bibitem [{\citenamefont {Phien}\ \emph {et~al.}(2012)\citenamefont {Phien},
  \citenamefont {Vidal},\ and\ \citenamefont {McCulloch}}]{phien2012infinite}%
  \BibitemOpen
  \bibfield  {author} {\bibinfo {author} {\bibfnamefont {H.~N.}\ \bibnamefont
  {Phien}}, \bibinfo {author} {\bibfnamefont {G.}~\bibnamefont {Vidal}}, \ and\
  \bibinfo {author} {\bibfnamefont {I.~P.}\ \bibnamefont {McCulloch}},\ }\href
  {\doibase 10.1103/PhysRevB.86.245107} {\bibfield  {journal} {\bibinfo
  {journal} {Phys. Rev.B}\ }\textbf {\bibinfo {volume} {86}},\ \bibinfo {pages}
  {245107} (\bibinfo {year} {2012})}\BibitemShut {NoStop}%
\bibitem [{\citenamefont {Ba{\~n}uls}\ \emph {et~al.}(2009)\citenamefont
  {Ba{\~n}uls}, \citenamefont {Hastings}, \citenamefont {Verstraete},\ and\
  \citenamefont {Cirac}}]{banuls2009matrix}%
  \BibitemOpen
  \bibfield  {author} {\bibinfo {author} {\bibfnamefont {M.~C.}\ \bibnamefont
  {Ba{\~n}uls}}, \bibinfo {author} {\bibfnamefont {M.~B.}\ \bibnamefont
  {Hastings}}, \bibinfo {author} {\bibfnamefont {F.}~\bibnamefont
  {Verstraete}}, \ and\ \bibinfo {author} {\bibfnamefont {J.~I.}\ \bibnamefont
  {Cirac}},\ }\href {\doibase 10.1103/PhysRevLett.102.240603} {\bibfield
  {journal} {\bibinfo  {journal} {Phys. Rev. Lett.}\ }\textbf {\bibinfo
  {volume} {102}},\ \bibinfo {pages} {240603} (\bibinfo {year}
  {2009})}\BibitemShut {NoStop}%
\bibitem [{\citenamefont {Mattis}(2006)}]{mattis2006theory}%
  \BibitemOpen
  \bibfield  {author} {\bibinfo {author} {\bibfnamefont {D.~C.}\ \bibnamefont
  {Mattis}},\ }\href@noop {} {\emph {\bibinfo {title} {The theory of magnetism
  made simple: an introduction to physical concepts and to some useful
  mathematical methods}}}\ (\bibinfo  {publisher} {World Scientific Publishing
  Company},\ \bibinfo {year} {2006})\BibitemShut {NoStop}%
\bibitem [{\citenamefont {Li}\ and\ \citenamefont
  {Koch}(2017)}]{li2017:inversion}%
  \BibitemOpen
  \bibfield  {author} {\bibinfo {author} {\bibfnamefont {A.~C.~Y.}\
  \bibnamefont {Li}}\ and\ \bibinfo {author} {\bibfnamefont {J.}~\bibnamefont
  {Koch}},\ }\href {\doibase 10.1088/1367-2630/aa8d5b} {\bibfield  {journal}
  {\bibinfo  {journal} {New Journal of Physics}\ }\textbf {\bibinfo {volume}
  {19}},\ \bibinfo {pages} {115010} (\bibinfo {year} {2017})}\BibitemShut
  {NoStop}%
\bibitem [{\citenamefont {Sachdev}(2007)}]{sachdev2007quantum}%
  \BibitemOpen
  \bibfield  {author} {\bibinfo {author} {\bibfnamefont {S.}~\bibnamefont
  {Sachdev}},\ }\href {\doibase 10.1017/CBO9780511973765} {\emph {\bibinfo
  {title} {Quantum phase transitions}}}\ (\bibinfo  {publisher} {Cambridge
  University Press},\ \bibinfo {address} {Cambridge},\ \bibinfo {year}
  {2007})\BibitemShut {NoStop}%
\bibitem [{\citenamefont {Scully}\ and\ \citenamefont
  {Zubairy}(1997)}]{scully1997ms}%
  \BibitemOpen
  \bibfield  {author} {\bibinfo {author} {\bibfnamefont {M.~O.}\ \bibnamefont
  {Scully}}\ and\ \bibinfo {author} {\bibfnamefont {M.~S.}\ \bibnamefont
  {Zubairy}},\ }\href@noop {} {\emph {\bibinfo {title} {Quantum Optics}}}\
  (\bibinfo {address} {Cambridge},\ \bibinfo {year} {1997})\BibitemShut
  {NoStop}%
\bibitem [{\citenamefont {Ford}\ and\ \citenamefont
  {O'Connell}(1996)}]{ford1996there}%
  \BibitemOpen
  \bibfield  {author} {\bibinfo {author} {\bibfnamefont {G.}~\bibnamefont
  {Ford}}\ and\ \bibinfo {author} {\bibfnamefont {R.}~\bibnamefont
  {O'Connell}},\ }\href {\doibase 10.1103/PhysRevLett.77.798} {\bibfield
  {journal} {\bibinfo  {journal} {Phys. Rev. Lett.}\ }\textbf {\bibinfo
  {volume} {77}},\ \bibinfo {pages} {798} (\bibinfo {year} {1996})}\BibitemShut
  {NoStop}%
\bibitem [{\citenamefont {Altland}\ and\ \citenamefont
  {Simons}(2010)}]{altland2010condensed}%
  \BibitemOpen
  \bibfield  {author} {\bibinfo {author} {\bibfnamefont {A.}~\bibnamefont
  {Altland}}\ and\ \bibinfo {author} {\bibfnamefont {B.~D.}\ \bibnamefont
  {Simons}},\ }\href {\doibase 10.1017/CBO9780511789984} {\emph {\bibinfo
  {title} {Condensed matter field theory}}}\ (\bibinfo  {publisher} {Cambridge
  University Press},\ \bibinfo {address} {Cambridge},\ \bibinfo {year}
  {2010})\BibitemShut {NoStop}%
\bibitem [{\citenamefont {Stoudenmire}\ and\ \citenamefont
  {White}(2010)}]{stoudenmire2010minimally}%
  \BibitemOpen
  \bibfield  {author} {\bibinfo {author} {\bibfnamefont {E.}~\bibnamefont
  {Stoudenmire}}\ and\ \bibinfo {author} {\bibfnamefont {S.~R.}\ \bibnamefont
  {White}},\ }\href {\doibase 10.1088/1367-2630/12/5/055026/meta} {\bibfield
  {journal} {\bibinfo  {journal} {New J. Phys.}\ }\textbf {\bibinfo {volume}
  {12}},\ \bibinfo {pages} {055026} (\bibinfo {year} {2010})}\BibitemShut
  {NoStop}%
\end{thebibliography}%

\newpage~\newpage~

\onecolumngrid
\section{Supplementary Material for: Fluorescence spectrum and thermalization in a driven coupled cavity array} 
\twocolumngrid

\section{Driven-Dissipative XY Model}

This section provides the derivation of the effective transverse field anisotropic XY model which we study, starting from a model of a coupled cavity array, following Refs.~\cite{bardyn2012majorana,joshi2013quantum}. 

We consider a 1D lattice of optical or microwave cavities supporting photon modes $b_j$ with tunneling amplitude $J$ between adjacent cavities, and an on-site optical nonlinearity $U$ which induces effective photon-photon interactions in each cavity. Such a coupled cavity array is thus described by the Bose-Hubbard Hamiltonian: 
$$
H = \sum_j \left[ \omega_c b^{\dag}_j b^{}_j + U b^{\dag}_j b^{\dag}_j b^{}_j b^{}_j  - J  \left( b^{\dag}_j b^{}_{j+1} + \text{H.c.} \right) \right].
$$
In addition to these elements, we consider a two-photon drive $\Omega \cos(2\omega_P t)$ near two-photon resonance $\omega_P \approx \omega_c$. We then work in the limit of strong optical nonlinearity with a perfect photon blockade, which restricts occupations to at most one photon in each cavity. This strong nonlinearity then implies that the two-photon drive is only resonant with creation of photon pairs on adjacent cavities. The above considerations allow us to replace each cavity mode with a spin-$1/2$, equivalent to replacing the bosonic operators by Pauli matrices: $b_j \to \sigma^{-}_j$. Our model Hamiltonian then becomes:
\begin{multline}
H_0 =  \sum_j \frac{\omega_c}{2} \sigma^z_j - J \sum_j \left[ \sigma^{+}_j \sigma^{-}_{j+1} + \text{H.c.} \right] \\ - \Omega \sum_j \left[ \sigma^{+}_j \sigma^{+}_{j+1} e^{-2i\omega_p t} + \text{H.c.}  \right].
\end{multline}
If we then define the dimensionless parameters $g=(\omega_p - \omega_c)/2J$, $\Delta=\Omega/J$, we can transform $H_0$ to a rotating frame (at pump frequency $\omega_p$) to gauge away the explicit time-dependence and write:
\begin{equation}
H =  -J \sum_j \left[ g \sigma^z_j + \frac{1+\Delta}{2} \sigma^x_j \sigma^x_{j+1} + \frac{1-\Delta}{2} \sigma^y_j \sigma^y_{j+1} \right].
\end{equation}
The Hamiltonian $H$ of a coupled cavity array thus takes a form of $XY$ model where $g$ acts as the transverse magnetic field, and $\Delta$ is the anisotropy of spin-spin interactions. The limit $\Delta=0$ corresponds to the isotropic XY model and $\Delta=1$ to the Ising model.

\section{Spin-wave approximation at small excitation number}

In this section we present further of the spin wave theory~\cite{mattis2006theory} used to describe the behavior at small excitation number.  In particular, we can use this to understand either the limit of large $|g|$ or small $\Delta$, as both lead to small excitation number. Such an approach was used in~\citet{joshi2013quantum} for small $\Delta$ to calculate static correlation functions; here we extend this to dynamical correlation functions and associated spectra.

At $\Delta = 0$ (i.e. zero pumping $\Omega/J = 0$), or at $g \to -\infty$, the NESS of our model corresponds to an empty state.  For a small $\Delta$, one can thus use spin-wave approximation, which ignores the constraint on double occupancy of a lattice site, and so is only valid for a low density of excitations.
 In this small excitation number regime we can revert from spin-$1/2$ operators (hard-core bosons) to bosonic fields: $\sigma_j^{-} \to b_j$, hence recovering aspects of a weakly interacting model. (Note that for large positive $g$ a similar argument can be made, making use of the duality under $g \to -g$ discussed in the manuscript.)

 \subsection{Calculating correlation functions}
\label{sec:calc-corr-funct}

\subsubsection{Spin-wave approximation and equations of motion}
\label{sec:spin-wave-appr}

We first follow the steps described in~\cite{joshi2013quantum} to derive the Hamiltonian in terms of Bosonic system operators $b_k$ and $b^{\dag}_{-k}$. Working in the momentum basis, $b_k =\sum_j e^{i k j} b_j/\sqrt{N}$, the master equation, written as Eq.~$\left(1 \right)$ in the main text becomes:
\begin{equation}
\partial_t \rho = -i \sum_k \left[h_{k},\rho \right] + \frac{\kappa}{2} \sum_k \left(2 \hat{b}_k \, \rho b^{\dag}_k - b^{\dag}_k b_k \, \rho - \rho \, b^{\dag}_k b_k \right),
\end{equation} 
where
\begin{equation}
h_k = - \left(\begin{matrix} b^{\dag}_k & b_{-k} \end{matrix} \right) \left(\begin{matrix} g + \cos(k) & \Delta \cos(k) \\ \Delta \cos(k) & g + \cos(k) \end{matrix} \right)  \left(\begin{matrix} b_k \\ b^{\dag}_{-k} \end{matrix} \right),
\end{equation}
and we have set $J=1$, so all energies are measured in units of $J$.
Note that when $\Delta$ controls the strength of pair creation,
while $\text{max}(\kappa,g,1)$ determines the cost of creating these excitations, so the small excitation regime corresponds to $\Delta \ll \text{max}(\kappa,g,1)$.  To find correlation functions, rather than considering the master equation above, we introduce the equivalent Heisenberg-Langevin equations for the system operators coupled to a Markovian bath~\cite{scully1997ms}. The equations of motion for $b^{}_k$ and $b^{\dag}_{-k}$ can be written in a matrix form:
\begin{equation} \label{Langevin}
\partial_t f(t) = M f(t) + v(t),
\end{equation}
with the vectors:
\begin{equation}
  f(t) = 
  \begin{pmatrix}
    b_k(t) \\ b^{\dag}_{-k}(t)
  \end{pmatrix},
  \qquad
  {v}(t) = 
  \sqrt{2 \kappa} 
  \begin{pmatrix} b^{\text{in}}_k(t) \\ b^{\dag\text{in}}_{-k}(t) \end{pmatrix},
\end{equation}
and the matrix:
\begin{equation} \label{dynamical-matrix}
M = \left(\begin{matrix} -\kappa + 2i(g + \cos(k)) & 2i\Delta \cos(k) \\ -2i\Delta \cos(k) & -\kappa - 2i(g + \cos(k)) \end{matrix} \right).
\end{equation}
Here, coupling to Markovian bath introduces the input noise term $b^{\text{in}}_k(t)$. Since we consider a zero temperature bath, there is only vacuum quantum noise, and the only nonzero correlator is $\left< b^{\text{in}}_k(t) b^{\dag\text{in}}_{k^{\prime}}(t^{\prime}) \right>  = \delta_{k,k^{\prime}} \delta(t-t^{\prime})$. 

The solution of \eqref{Langevin} is: 
$$f(t) = e^{Mt} f(0) + \int^{t}_0 dt^{\prime} e^{M(t-t^{\prime})} v(t^{\prime}).$$ 
In the long-time limit $t \to \infty$ we find the expressions for system operators:
\begin{align} \label{operator-sol}
b_k(t) &= \sqrt{2 \kappa} \int^{t}_0 \!\! dt^{\prime} \left[ G_1(t-t^{\prime}) b^{\text{in}}_k(t^{\prime})  +  G_2(t-t^{\prime}) b^{\dag\text{in}}_{-k}(t^{\prime}) \right],
\nonumber\\
{b}^{\dag}_{-k}(t) &= \sqrt{2 \kappa} \int^{t}_0 \!\! dt^{\prime} \left[ G^{*}_1(t-t^{\prime}) b^{\dag\text{in}}_{-k}(t^{\prime})  +  G^{*}_2(t-t^{\prime}) b^{\text{in}}_k(t^{\prime}) \right].
\end{align}
where the propagators $G_{1,2}(\tau)$ are matrix elements of $e^{Mt}$ given by:
\begin{align}
G_1(\tau) &= e^{-\kappa \tau} \left[ \cos(\xi_k \tau) + i \epsilon_k \frac{\sin(\xi_k \tau) }{\xi_k} \right],
\\
G_2(\tau) &= i \eta_k e^{-\kappa \tau} \frac{\sin(\xi_k \tau) }{\xi_k},
\end{align}
with dispersions $\epsilon_k = 2(g + \cos(k))$, $\eta_k = 2\Delta\cos(k)$, and $\xi_k = \sqrt{\epsilon^2_k - \eta^2_k}$.

\subsubsection{Correlations and effective temperatures for $\hat\sigma_x$}
\label{sec:corr-effect-temp}

After deriving the system operators, we now proceed to calculate the frequency-resolved spectra for $XX$ correlations. Since $\sigma^x_j \to b_j + b^{\dag}_j$ in the spin-wave limit, we can express the on-site $XX$ two-time correlator as:
\begin{multline} \label{Cxx-orig}
\tilde{C}^{xx}(\tau) = \left< \sigma^x(0) \sigma^x(\tau) \right> = \left< b^{\dag}(0) b^{\dag}(\tau) \right> + \left< b(0) b(\tau) \right> \\ + \left< b^{\dag}(0) b(\tau) \right> + \left< b(0) b^{\dag}(\tau) \right>,
\end{multline}
where the correlations are given by a Fourier transform from momentum to real space: 
$$
\left< b^{\dag}(0) b(\tau) \right> = \int^{\pi}_{-\pi} \; dk \; e^{ikl}  \left< b^{\dag}_k(0) b_k(\tau) \right> \bigg\rvert_{l=0},
$$ 
and similar expressions for other correlators. We then substitute in the solutions for operators \eqref{operator-sol}, and evaluate the time integrals at unequal times, $t + \tau$ and $t$. This gives a two-time correlator:
\begin{multline} \label{Cxx-final}
\tilde{C}^{xx}(\tau) = \frac{e^{-\kappa \tau}}{2\pi} \int^{\pi}_{-\pi} \!\! dk \;
\Biggl[  \cos(\xi_k \tau) + i(\eta_k - \epsilon_k) \frac{\sin(\xi_k \tau) }{\xi_k} \\ +  
\frac{\eta_k (\eta_k - \epsilon_k)}{\xi_k^2 + \kappa^2}  \left(\cos(\xi_k \tau) + \kappa \frac{\sin(\xi_k \tau)}{\xi_k}  \right) \Biggr].
\end{multline}

The quantities of interest are the fluctuation spectrum $S(\omega)$ and susceptibility $\chi^{\prime\prime}(\omega)$ given by the Fourier transforms of $\tilde{S}(\tau) = \frac{1}{2} (\tilde{C}(\tau)^* + \tilde{C}(\tau))$ and $\tilde{\chi}(\tau) = i \Theta(\tau) (\tilde{C}(\tau)^* - \tilde{C}(\tau))$ respectively. 
Plugging in \eqref{Cxx-final} and taking a Fourier transform with respect to $\tau$, we obtain $XX$ spectra:
\begin{align}
  \label{Sxx}
  S_{xx}(\omega) &= \frac{\kappa}{\pi} \int^{\pi}_{-\pi} dk \; \frac{P_k + \omega^2 +2\eta_k (\eta_k - \epsilon_k) }{Q_k(\omega)},
  \\
 \label{Chixx}
\chi^{\prime\prime}_{xx}(\omega) &= \frac{2\kappa\omega}{\pi} \int^{\pi}_{-\pi} dk \;  \frac{\eta_k - \epsilon_k}{Q^{-1}_k(\omega)},
\end{align}
where we have introduced auxiliary functions $P_k = \xi_k^2 + \kappa^2$, and $Q_k(\omega) = (P_k - \omega^2)^2 + (2\omega\kappa)^2$. 

One can then substitute $e^{ik} \to z$, and the integrals in \eqref{Sxx}, \eqref{Chixx} become contour integrals around a unit circle $C$ with $|z| = 1$. The values of \eqref{Sxx}, \eqref{Chixx} are then determined by the residues of poles $z=Z$ located inside $C$ (i.e. with $|Z| < 1$). Both \eqref{Sxx} and \eqref{Chixx} have the same set of eight poles given by:
\begin{equation*}
Z = \theta_{s_1, s_2} \pm \sqrt{\theta^2_{s_1, s_2} - 1},
\end{equation*}
\begin{equation*}
\theta_{s_1, s_2}= \frac{ -g + s_1 \sqrt{g^2 \Delta^2 - \frac{1 - \Delta^2}{4}\left[\kappa^2 - \omega^2 + s_2 \, 2i\omega\kappa \right]}}{1 - \Delta^2},
\end{equation*}
with $s_1 = \pm 1$, $s_2 = \pm 1$. Evaluating the contour integrals gives fluctuation spectrum:
\begin{equation} \label{Sxx-result}
S_{xx}(\omega) = 2\kappa\sum_{|Z_n|<1} Z_n \alpha_n,
\end{equation}
where
\begin{equation*}
\alpha_n = \frac{\left[(1-\Delta)Z_n^2 + 2 g Z_n + (1-\Delta)\right]^2 + (\omega^2 + \kappa^2)Z_n^2}{(1-\Delta^2)^2 \prod^{8}_{m=1, m \neq n} (Z_n - Z_m)}.
\end{equation*}
Similarly the  susceptibility is given by:
\begin{equation} \label{Chixx-result}
\chi^{\prime\prime}_{xx}(\omega) = -4\kappa\omega \sum_{|Z_n|<1} Z_n^2 \beta_n,
\end{equation}
where
\begin{equation*}
\beta_n = \frac{(1-\Delta)Z_n^2 + 2 g Z_n + (1-\Delta)}{(1-\Delta^2)^2 \prod^{8}_{m=1, m \neq n} (Z_n - Z_m)}.
\end{equation*}
In both expressions, the sum runs over poles $Z_n$ inside the unit circle $C$ with $|Z_n| < 1$. From \eqref{Sxx-result} and \eqref{Chixx-result} it is straightforward to derive the distribution function $F_{xx}(\omega)$ of fluctuation-dissipation theorem:
\begin{equation}
F_{xx}(\omega) = \frac{S_{xx}(\omega)}{\chi^{\prime\prime}_{xx}(\omega)} = - \frac{1}{2 \omega} \; \frac{\sum_{|Z_n|<1} Z_n \alpha_n}{\sum_{|Z_n|<1} Z_n^2 \beta_n}.
\end{equation}
It is evident that in the low frequency limit $\omega \to 0$, the distribution $F_{xx}(\omega)$ is dominated by the ${1}/{\omega}$ divergence. The effective thermalization of NESS thus already emerges in spin-wave theory, leading to an effective temperature $T_{\text{eff}}$:
\begin{equation}
T_{\text{eff}, xx} = - \frac{1}{4} \; \frac{\sum_{|Z_n|<1} Z_n \alpha_n}{\sum_{|Z_n|<1} Z_n^2 \beta_n} \bigg\rvert_{\omega=0}.
\end{equation}

\subsubsection{Correlations and effective temperatures for $\hat\sigma_y$}

Similarly to the previous subsection, we can calculate the fluctuation spectrum and susceptibility for $YY$ excitations in the spin-wave limit where $\sigma_j^y \to -i(b_j - b_j^{\dag})$. By following the same steps as for $XX$ correlators, we obtain the $YY$ spectra:
\begin{equation} \label{Syy}
S_{yy}(\omega) = \frac{\kappa}{\pi} \int^{\pi}_{-\pi} dk \; Q^{-1}_k(\omega) \left[P_k + \omega^2 +2\eta_k (\eta_k + \epsilon_k) \right],
\end{equation}
\begin{equation} \label{Chiyy}
\chi^{\prime\prime}_{yy}(\omega) = \frac{2\kappa\omega}{\pi} \int^{\pi}_{-\pi} dk \; Q^{-1}_k(\omega) (\eta_k + \epsilon_k).
\end{equation}
The only differ from the expressions for $XX$ spectra \eqref{Sxx}, \eqref{Chixx} by $(\eta_k - \epsilon_k) \to (\eta_k + \epsilon_k)$.  Note that as a result,
the contours integrals to evaluate have the same poles as for $XX$ correlators, but with different residues and thus different weights.   

Continuing along the same steps as for $XX$ correlators, we find the $YY$ fluctuation spectrum:
\begin{equation} \label{Syy-result}
S_{yy}(\omega) = 2\kappa\sum_{|Z_n|<1} Z_n \gamma_n,
\end{equation}
where
\begin{equation*}
\gamma_n = \frac{\left[(1+\Delta)Z_n^2 + 2 g Z_n + (1+\Delta)\right]^2 + (\omega^2 + \kappa^2)Z_n^2}{(1-\Delta^2)^2 \prod^{8}_{m=1, m \neq n} (Z_n - Z_m)}.
\end{equation*}
The $YY$ susceptibility is given by:
\begin{equation} \label{Chiyy-result}
\chi^{\prime\prime}_{yy}(\omega) = -4\kappa\omega \sum_{|Z_n|<1} Z_n^2 \delta_n,
\end{equation}
where
\begin{equation*}
\delta_n = \frac{(1+\Delta)Z_n^2 + 2 g Z_n + (1+\Delta)}{(1-\Delta^2)^2 \prod^{8}_{m=1, m \neq n} (Z_n - Z_m)}.
\end{equation*}
and the poles $Z_n$ are the same as in $XX$ spectra. One can then straightforwardly derive the distribution function $F_{yy}(\omega)$:
\begin{equation}
F_{yy}(\omega) = \frac{S_{yy}(\omega)}{\chi^{\prime\prime}_{yy}(\omega)} = - \frac{1}{2 \omega} \; \frac{\sum_{|Z_n|<1} Z_n \gamma_n}{\sum_{|Z_n|<1} Z_n^2 \delta_n}.
\end{equation}
and the effective temperature $T_{\text{eff}}$:
\begin{equation}
T_{\text{eff}, yy} = - \frac{1}{4} \; \frac{\sum_{|Z_n|<1} Z_n \gamma_n}{\sum_{|Z_n|<1} Z_n^2 \delta_n} \bigg\rvert_{\omega=0}.
\end{equation}

\subsubsection{Vanishing correlations for $\sigma_z$}

Note that while the above approach allows calculation of the $XX$ and $YY$ correlators, we cannot use this small excitation number approximation to find
the $ZZ$ correlators.  To see this, note that if we express the $ZZ$ two-time correlator using $\sigma^z_j =  b^{\dag}_j b_j - b_j b^{\dag}_j$, then in the spin-wave limit:
\begin{multline} \label{Czz-orig}
\tilde{C}^{zz}(\tau) = \left< \sigma^z(0) \sigma^z(\tau) \right> = \\ = \left< b^{\dag}(0) b(0) b^{\dag}(\tau) b(\tau) \right> - \left< b(0) b^{\dag}(0) b^{\dag}(\tau) b(\tau) \right> \\ + \left< b(0) b^{\dag}(0) b(\tau) b^{\dag}(\tau) \right> - \left< b^{\dag}(0) b(0) b(\tau) b^{\dag}(\tau) \right>.
\end{multline}
Since the problem involves non-interacting bosons, the steady state is Gaussian and we can expand the four-field correlators using Wick's theorem, which leads to $\tilde{C}^{zz}(\tau) = 0$.  Thus, at this order of approximation
all $ZZ$ spectra are trivially zero.  This is to be expected, since $\sigma^z$  correlations are quartic and the spin wave theory is linear.

\subsection{Temperature of individual bosonic modes}

Building on the spin-wave theory introduced above, we next discuss thermalization and the appearance of an effective temperature from this linear theory.  To understand these effects, it is helpful to consider the contribution of individual bosonic modes. From \eqref{Sxx}, \eqref{Chixx} we thus define the momentum-resolved fluctuation spectrum and susceptibility: 
\begin{equation} \label{Sxx_kk}
S_{xx}(\omega, k) = \frac{\kappa}{\pi} \; Q^{-1}_k(\omega) \left[P_k + \omega^2 +2\eta_k (\eta_k - \epsilon_k) \right].
\end{equation}
\begin{equation} \label{Chixx_kk}
\chi^{\prime\prime}_{xx}(\omega, k) = \frac{2\kappa\omega}{\pi} \; Q^{-1}_k(\omega) (\eta_k - \epsilon_k).
\end{equation}
Then the distribution function of an individual bosonic $k$-mode:
\begin{equation} \label{Fxx_kk}
F_{xx}(\omega, k) = \frac{S_{xx}(\omega, k)}{\chi^{\prime\prime}_{xx}(\omega, k)} 
= \frac{\xi_k^2 + \omega^2 + \kappa^2 + 2 \eta_k (\eta_k - \epsilon_k)}{2 \omega (\eta_k - \epsilon_k)}.
\end{equation}
Focusing on the frequency dependence of this expression, we see it can be written in the form:
\begin{equation} \label{Fxx_kk_result}
F_{xx}(\omega, k) = \frac{2 T_{\text{eff},xx,k} + \lambda_{xx,k} \omega^2}{\omega},
\end{equation}
where $\lambda_{xx,k} = [2(\eta_k - \epsilon_k)]^{-1}$, and the effective temperature of an individual bosonic mode, defined from the $\omega \to 0$ limit is:
\begin{equation} \label{Teff_kk}
T_{\text{eff},xx,k} = \frac{\kappa^2 +  (\eta_k - \epsilon_k)^2}{4 (\eta_k - \epsilon_k)},
\end{equation}
where we have used the definition of $\xi_k^2 = \epsilon_k^2 - \eta_k^2$.

From the above, we see that despite considering a linearized (i.e. non-interacting) theory, a low energy effective temperature emerges for each individual $k$ mode.   However, the functional form of $F_{xx}(\omega, k)$ does not show a plateau around $F_{xx}(\omega, k) \simeq 1$, as expected for an equilibrium system, and as sometimes seen from the MPS numerics for $F_{xx}(\omega)$.  Instead $|F_{xx}(\omega, k)|^{-1}$ shows a peak
at  $\omega = \omega^\ast_{xx,k} \equiv \sqrt{2 T_{\text{eff},xx,k}/\lambda_{xx,k}}=
\sqrt{\kappa^2 + (\eta_k - \epsilon_k)^2}$, with a peak height
\begin{displaymath}
  |F_{xx}(\omega^\ast_{xx,k}, k)|^{-1}
  =  \frac{1}{\sqrt{8 T_{\text{eff},xx,k}\lambda_{xx,k}}}
  = \sqrt{
\frac{(\eta_k - \epsilon_k)^2}{\kappa^2 + (\eta_k - \epsilon_k)^2}}.
\end{displaymath}
One may see that as required, this peak value is always less than one, and approaches one if damping is weak compared the energy difference $|\eta_k -\epsilon_k|$.  While this momentum resolved distribution function does not show a plateau, as seen in Fig.~2, a plateau does arise for the site-local
(i.e. momentum integrated) result.  This site-local distribution function can be expressed as a weighted average of distributions $F(\omega, k)$ of individual bosonic modes:
\begin{equation}\label{eq:Fxx_sum}
F(\omega) = \frac{\int_{-\pi}^{\pi} dk \; F(\omega, k) \chi^{\prime\prime}(\omega, k) }{\int_{-\pi}^{\pi} dk \; \chi^{\prime\prime}(\omega, k)}.
\end{equation}
Because the location of the peak for each mode $k$ differs, this weighted average shows a plateau arising from combining all these peaks, stretching over a range of frequencies, set by the range of peak frequencies, i.e. $\sqrt{\kappa^2  + 4[g-(1-\Delta)]^2}<\omega^\ast_{xx,k}<\sqrt{\kappa^2  + 4[g-(1+\Delta)]^2}$.  Note that as such the location of the plateau moves to higher frequencies as we increase $g$, as seen in Fig.~(2) of the ma text.  Note also that the plateau is always finite, and $F(\omega) \propto \omega$ at large enough frequency.

As noted earlier, if we look at $YY$ correlations in place of $XX$, the only change is to replace $\eta_k-\epsilon_k \to \eta_k + \epsilon_k$ in the above expressions, including in the density of states $\chi^{\prime\prime}(\omega, k)$.  Because of this change, one has both that $T_{\text{eff},yy,k}$ differs from
$T_{\text{eff},xx,k}$, as well as the distribution of occupied modes changing.

In the limit of large $g$, Eq.~(\ref{Teff_kk}) becomes
$T_{\text{eff},xx,k} \approx T_{\text{eff},yy,k} \approx -g/2$, independent of momentum $k$ and the operator being measured. Since this result is independent of momentum, the local effective temperature from Eq.~(\ref{eq:Fxx_sum}) also approaches this value, 
\begin{equation}
T_{\text{eff}} \approx -g/2.
\end{equation}
In Fig.(3) of the main text we show the extracted temperature of the spin-wave theory for both $\sigma_x$ and $\sigma_y$ correlators:  at large
$g$, where quasi-thermalization holds, we see both temperatures approach this
same value.  In the opposite limit, of small $g$, one may note that
$\eta_k \pm \epsilon_k = 2(\Delta \pm 1) \cos(k) \pm g$ can now pass through zero for some real $k$, leading to a divergence of both $T_{\text{eff},k}$ and $\lambda_k$.  This divergence is however integrable, giving a finite form of $F(\omega)$ and the corresponding $T_{\text{eff}}$, except as seen at $g=0$
in Fig.(3).

\subsection{Correlations for the Ising limit}
\label{sec:corr-ising-limit}

\begin{figure}[htpb]
  \centering
  \includegraphics[width=\linewidth]{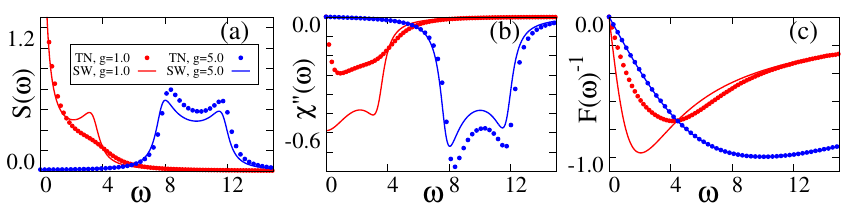}
  \caption{Correlation functions for $\Delta=1$, comparing MPS numerics (points) with spin wave calculations (lines).  Panels (a-c) show the spectrum of fluctuations $S(\omega)$, imaginary part of response function $\chi^{\prime\prime}(\omega)$, and the inverse distribution function $F(\omega)^{-1}$ respectively.  The results match well for $g=5.0$, but poorly for $g=1.0$.}
  \label{fig:delta1-sw}
\end{figure}

As noted earlier, the small excitation density limit can be understood as resulting either from small $\Delta$ or large $g$.  As such, this approximation should remain valid even for the Ising limit,  $\Delta=1$, as long as $g \gg 1$.  However, at first appearance the above results are singular in the limit $\Delta=1$.  This is in fact not the case as one may readily check.  For example, considering $\hat O = \hat \sigma^x$, we see that as $\Delta \to 1$ the poles
given $Z$ are still given by $\theta_{s_1,s_2}$ but the values of $\theta_{s_1,s_2}$ become singular, specifically:
\begin{displaymath}
  \theta_{s_1,s_2} =
  \begin{cases}
    - (\kappa + s_2 i \omega)^2 / 8g & s_1 = +1 \\
    - {g}/(1-\Delta) & s_1=-1
  \end{cases}.
\end{displaymath}
Inserting this into the definition of $Z$, we find that of the eight poles,
four remain finite, while two tend to zero and two to infinity.  Pairs of
poles at zero and infinity in fact cancel, as long as the residue at the remaining poles is finite.  This can be checked to be true, with the residues
becoming $\alpha_n=(\omega^2 + \kappa^2)/16 g^2$ for the finite poles.  The behavior in the limit $\Delta \to 1$ is shown in Fig.~\ref{fig:delta1-sw}.

As seen in Fig.~\ref{fig:delta1-sw}, the spin wave theory indeed accurately captures the behavior at large $g$, but clearly fails in the case $g=1.0$, where it would not be expected to hold.  As with the small $\Delta$ results in the main text, we see that the distribution function matches more accurately than the fluctuation and response functions separately.  We may note that for $\Delta=1$, there is never a true plateau in the distribution function.  This can be understood from the fact that for $\Delta=1$, $\epsilon_k -\eta_k= 2g$, so both $T_{\text{eff},xx,k}=-(\kappa^2 + 4 g^2)/4g$ and $\lambda_{xx,k}=-1/2g$ become independent of $k$, meaning that $F_{xx}(\omega,k)$ is also independent of $k$, so the integrated version $F(\omega)$ follows the form of Eq.~(\ref{Fxx_kk_result}), having a peak at $\omega^\ast = \sqrt{\kappa^2 + 4g^2}$ as is visible in Fig.~\ref{fig:delta1-sw}.  Note however that for $YY$ correlations the same statement would not be true, as $\epsilon_k + \eta_k= 2g + 4 \cos(k)$ is $k$-dependent.

\subsection{Fluctuation-dissipation relation in linear theories}

It is notable that, as discussed above, the linearized the spin-wave result predicts a low energy quasi-thermal distribution with a non-zero effective temperature for $XX$ and $YY$ excitations.  This is particularly notable in the light of papers, e.g.~\citet{ford1996there} which suggest that the fluctuation dissipation theorem fails for a Markovian dissipation of a bath, and one would expect to find the effective $F(\omega)$ to be frequency independent, corresponding approximately to zero temperature. This section discusses why the model we consider does not show such behavior.  Specifically, as already noted, the structure of distribution function is dependent on the mode considered.  If in place of the $XX$ and $YY$ correlations we had directly considered correlations of the anihilation operators, $\tilde{C}_{bb^\dagger}(\tau) = \left< b(0) b^\dagger(\tau) \right>$, then we would have found the distribution function $F_{bb^\dagger}(\omega)$ would be flat.  We first discuss this point, and why it is that $XX$ and $YY$ distribution functions are not flat.  Since Ref.~\cite{ford1996there} also considers the analogue of $XX$ correlations, we then address further differences between the model discussed there and our results.

We first discuss the observation that the distribution function for correlators of annihilation and creation operators generally leads to a flat distribution. A quantum harmonic oscillator (with frequency $\Omega$, and field operators $b$, $b^{\dag}$) interacting with a bath of radiation modes (with frequency $\omega_k$, field operators $B_k^{\dag}$, $B_k$ for each mode) via coupling strength  is described by the Hamiltonian
\begin{equation}
H = \Omega b^{\dag} b + \sum_k \omega_k B_k^{\dag} B_k + \sum_k \left[g_k b B_k^{\dag} + \text{H.c.}\right].
\end{equation}
Here $g_k$ is the system-bath coupling strength. Using the standard input-output formalism~\cite{scully1997ms} in the Markovian limit we derive Heisenberg-Langevin equation of motion
\begin{equation}
\partial_t b(t) = -i \Omega b(t) - \kappa b(t) + \sqrt{2 \kappa} b^{\text{in}}(t),
\end{equation}
where the input noise operator $b^{\text{in}}(t)$ is introduced by coupling to Markovian bath. For a zero-temperature bath there is only vacuum noise and the only non-zero correlator is $\left< b^{\text{in}}(t) b^{\dag\text{in}}(t^{\prime}) \right> = \delta(t-t^{\prime})$. Next, one can obtain the steady-state solution (at $t \to \infty$): 
\begin{equation}
b(t) = \sqrt{2 \kappa} \int_{-\infty}^{t} dt^{\prime} e^{-i\Omega(t-t^{\prime}) - \kappa(t-t^{\prime})} b^{\text{in}}(t).
\end{equation}
The only non-vanishing two-time correlator then is
\begin{equation}
\langle b(0) b^{\dag} (\tau) \rangle = e^{-\kappa |\tau| + i\Omega\tau}.
\end{equation}
Taking a Fourier transform of the symmetrized correlator $\tilde{S}_{bb^\dagger}(\tau) = \frac{1}{2}\left<\lbrace b(0), b^{\dag}(\tau)\rbrace \right> = e^{-\kappa |\tau| + i\Omega\tau}$ and response function $\tilde{\chi}_{bb^\dagger}(\tau) = i \theta(\tau) \left<\left[b(0), b^{\dag}(\tau) \right] \right> = i \theta(\tau) \, e^{-\kappa |\tau| + i\Omega\tau} $ gives the fluctuation spectrum and susceptibility:
\begin{equation}
S_{bb^\dagger}(\omega) = \chi^{\prime\prime}_{bb^\dagger}(\omega) = \frac{2\kappa}{(\omega - \Omega)^2 + \kappa^2}.
\end{equation}
Subsequently, one obtains a flat distribution spectrum for $b$, $b^{\dag}$ modes, in contrast to the quasi-thermal distribution of the $XX$ and $YY$ modes:
\begin{equation}
F_{bb^\dagger}(\omega) = \frac{S_{bb^\dagger}(\omega)}{\chi^{\prime\prime}_{bb^\dagger}(\omega)} = 1.
\end{equation}

Our spin-wave equations differs from the above derivation in that the spin-wave theory has anomalous terms proportional to $\Delta$.  However, the derivation above can be extended just as well to a linear theory with anomalous terms since its Hamiltonian can be diagonalized easily using the Bogoliubov transformation~\cite{altland2010condensed}.  The crucial difference that occurs in the spin wave theory discussed above is our calculation of $XX$ and $YY$ correlations, which mean fluctuation and dissipation terms involve sums and differences of correlators $\langle b(0) b^\dagger(\tau)\rangle$ and $\langle b^\dagger(0) b^{}(\tau)\rangle$.  Once these are both considered, the single mode functions results in Eq.~(\ref{Sxx_kk}--\ref{Fxx_kk}) follow, giving a frequency dependent result.

As noted earlier, the above result is notable in connection to the argument by~\citet{ford1996there} that quantum regression can never give a thermal spectrum. The problem considered there is similar to ours in that there are anomalous terms (since no rotating-wave approximation is made in the system bath coupling), and the correlations considered are the $XX$ correlations.  However, there is a crucial difference in that Ref.~\cite{ford1996there} considers Ohmic  rather than Markovian dissipation.  This difference leads to the different conclusions of the previous section.

\section{Tensor network approach for two-time correlations}

\begin{figure} 
  \begin{center}
\includegraphics[width=1.0\linewidth]{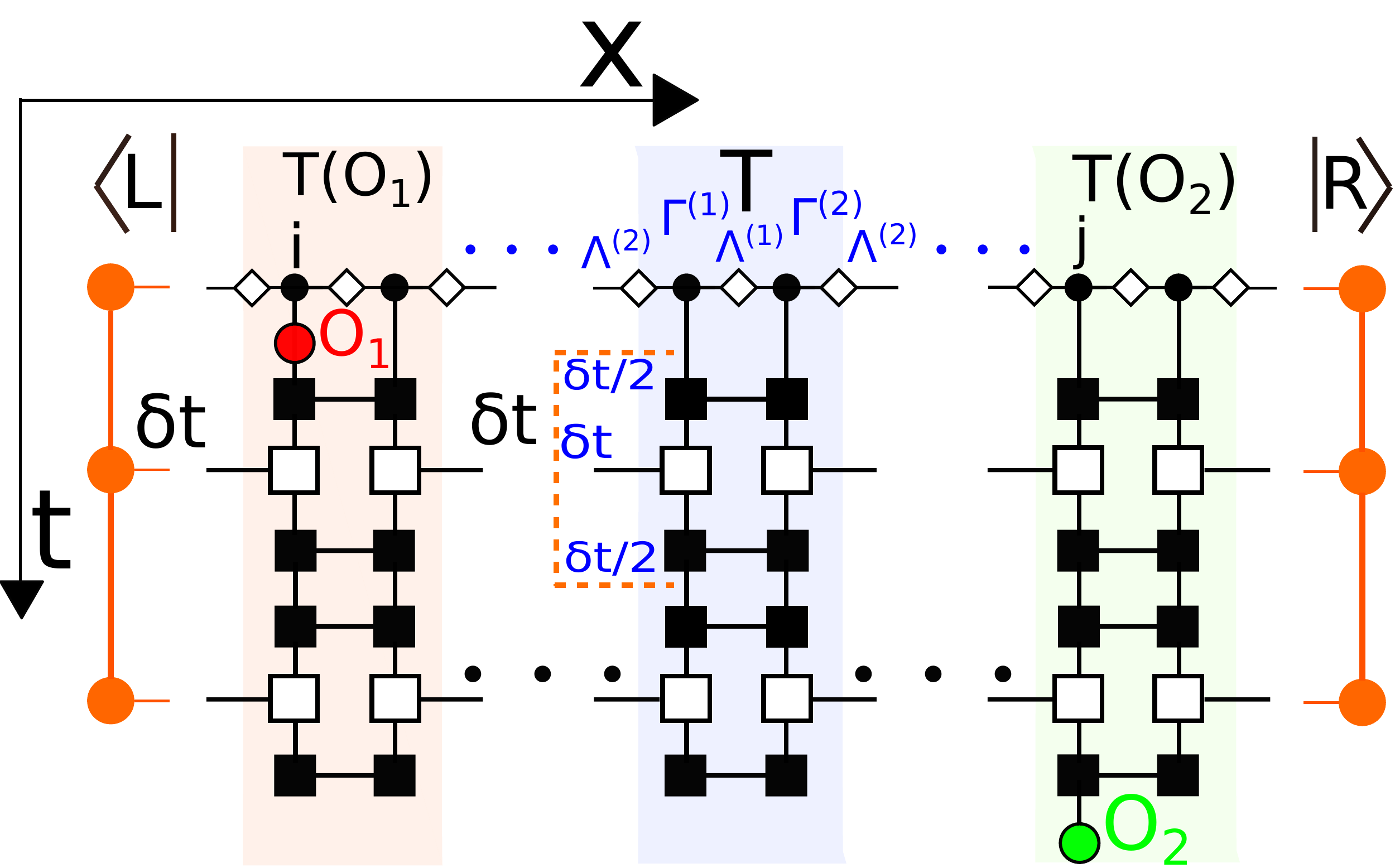}
  \end{center}
\caption{Tensor network used to evaluate two-time correlations, using boundary eigenvectors (orange) for calculating two-time correlations under non-unitary Liouvillian propagator.}
\label{CCA-Cartoon}
\end{figure}

In this section, we describe the tensor network method for computing two-time correlations in open quantum systems in the thermodynamic limit. We use quantum regression to calculate two-time correlations Eq.~\eqref{Q-reg}, starting from the NESS density matrix $\rho_{SS}$, represented by a translationally invariant infinite MPS that was previously computed using infinite TEBD algorithm~\cite{schollwock2011density,orus2014practical}. 
\begin{equation} \label{Q-reg}
\left< O_2^{(j)}(t) O_1^{(i)}(0) \right> = \Tr\left[O_2^{(j)} \, e^{t \mathcal{L}} \, O_1^{(i)}\rho_{\text{ss}}\right].
\end{equation}
For a finite size lattice, one could still directly use TEBD algorithm to perform the time evolution in \eqref{Q-reg}. However, such direct propagation is incompatible with infinite TEBD: application of a local operator $O_1$ to $\rho_{SS}$ breaks translational invariance. A naive solution would be to use a finite size extrapolation, which is prone to boundary and finite-size effects. In particular, the finite lattice size would restrict the extent of correlations in both space and time, as excitations will be reflected back from the boundaries, and the simulation will be no longer valid at later times~\cite{phien2012infinite}. Such simulation would also inccur an additional computational cost that scales linearly with the system size which will be inefficient for large lattices needed to approximate the thermodynamic limit, in comparison to only two sites required in the infinite TEBD.  Nonetheless, such a method has been recently used to calculate aging dynamics in the XXZ model~\cite{wolff2018evolution}.

Fortunately, it is possible to avoid these issues that arise due to  finite size altogether. A method to compute two-time correlations in an infinite system directly (i.e. without resorting to a finite size extrapolation) has been proposed by Ba$\tilde{\text{n}}$uls et al~\cite{banuls2009matrix} for unitary evolution in isolated systems. In our work, we extend this approach to open quantum systems whose dynamics is governed by the quantum master equation for density matrices. We provide a brief description of the algorithm below.

The idea is to construct a network representing the entire time evolution, instead of evolving the MPS in time step by step. We start from an infinite MPS representing vectorized NESS density matrix  $|\rho_{\text{ss}} \rangle$ and apply the first operator $O_1^{(i)}(0)$ at the initial time $t=0$. Then, for every time evolution step we insert a propagator MPO. After repeating this for the required number of time steps we apply the second operator $O_2^{(j)}(t)$ at the final time $t$. Taking the trace at the final time removes the dangling physical dimension at each site of the last MPO propagator. This procedure produces a 2D tensor network that is infinite along the spatial axis but finite along the time axis, giving an unnormalized two-time two-point correlator: 

\begin{equation}
\left< O_2^{(j)}(t)O_1^{(i)}(0) \right> \; \propto \; T^{N \to \infty} \; T_{O_1} \; T^{|i-j|-1} \; T_{O_2} \; T^{N \to \infty},
\end{equation}
where $T = $ is a transfer matrix of the evolved density matrix, and $T_{O_{1,2}}$ are transfer matrices containing an application of operators $O_{1,2}$ at the initial and final times at lattice sites $i$, $j$, as shown in Fig. 1(b) of the main text.

Since the network is translationally invariant, $T$ is the same on every site (except at the sites where the operators are applied) and we may effectively replace the semi-infinite lattices, to the left and right of the sites where $\hat{O}_1$ and $\hat{O}_2$ act, by the left and right eigenvectors of $T$ corresponding to its largest eigenvalue $\lambda$ since $\lim_{N \rightarrow \infty} T^N = \lambda^N |R \rangle \langle L|$. In practice, we compute an MPS approximation to the eigenvectors $|R \rangle$, $\langle L|$ by using the MPS-MPO power method. We multiply an initial arbitrary MPS (oriented along the time axis) by $T$ (represented as an MPO along the time axis) a sufficient number of times until it converges to $|R \rangle$, $\langle L|$ for the right- and left-multiplication respectively. We truncate the MPS bonds after each multiplication using the method described in~\cite{stoudenmire2010minimally}, performing truncation along the time axis.
Once we have calculated $|R \rangle$ and $\langle L|$, the resulting network is finite along both space
and time axes. It can then be easily contracted using MPO-MPS and MPS-MPS multiplications~\cite{schollwock2011density,stoudenmire2010minimally} to give any two-time two-point correlator: 
\begin{equation}
\left< O_2^{(j)}(t)O_1^{(i)}(0) \right> = \frac{\left<L| T_{O_1} T^{|i-j|-1} T_{O_2} |R \right>}{ \lambda^{|i-j|+1}},
\end{equation}
normalized by trace $\Tr(\rho) = \left<L| \, T^{|i-j|+1} |R \right> = \lambda^{|i-j|+1} $.

\end{document}